\documentclass[9pt,twocolumn,twoside]{pnas-new}

\usepackage{glossaries}
\usepackage{bm}

\newacronym{tr-RIXS}{tr-RIXS}{time-resolved resonant inelastic x-ray scattering}
\newacronym{tr-REXS}{tr-REXS}{time-resolved resonant elastic x-ray scattering}

\templatetype{pnasresearcharticle} 

\title{Laser-Induced Transient Magnons in Sr$_3$Ir$_2$O$_7$ Throughout the Brillouin Zone}

\author[a, b, 1]{Daniel G. Mazzone}
\author[a, c, 1]{Derek Meyers}
\author[a, d]{Yue Cao}
\author[e]{James G. Vale}
\author[e]{Cameron D. Dashwood}
\author[f]{Youguo Shi}
\author[g]{Andrew J. A. James}
\author[h]{Neil J. Robinson}
\author[a, i]{Jiaqi Lin}
\author[j]{Vivek Thampy}
\author[k]{Yoshikazu Tanaka}
\author[l]{Allan S. Johnson}
\author[a]{Hu Miao}
\author[i]{Ruitang Wang}
\author[a]{Tadesse A. Assefa}
\author[m]{Jungho Kim}
\author[m]{Diego Casa}
\author[n]{Roman Mankowsky}
\author[o]{Diling Zhu}
\author[o]{Roberto Alonso-Mori}
\author[o]{Sanghoon Song}
\author[o]{Hasan Yavas}
\author[k]{Tetsuo Katayama}
\author[k]{Makina Yabashi}
\author[p]{Yuya Kubota}
\author[p]{Shigeki Owada}
\author[q]{Jian Liu}
\author[q]{Junji Yang}
\author[a]{Robert M. Konik}
\author[a, e]{Ian K. Robinson}
\author[r]{John P.  Hill}
\author[e]{Desmond F. McMorrow}
\author[n]{Michael F\"orst}
\author[l, s, 2]{Simon Wall}
\author[i, 2]{Xuerong Liu}
\author[a, 2]{Mark P. M. Dean}

\affil[a]{Condensed Matter Physics and Materials Science Department, Brookhaven National Laboratory, Upton, New York 11973, USA}
\affil[b]{Laboratory for Neutron Scattering and Imaging, Paul Scherrer Institut, CH-5232 Villigen, Switzerland}
\affil[c]{Department of Physics, Oklahoma State University, Stillwater, Oklahoma 74078, USA}
\affil[d]{Materials Science Division, Argonne National Laboratory, Argonne, IL 60439, USA}
\affil[e]{London Centre for Nanotechnology and Department of Physics and Astronomy, University College
London, London WC1E 6BT, UK}
\affil[f]{Beijing National Laboratory for Condensed Matter Physics, Institute of Physics, Chinese Academy of Sciences, Beijing 100190, China}
\affil[g]{School of Physical Sciences, The Open University, Milton Keynes, MK7 6AA, UK}
\affil[h]{Institute for Theoretical Physics, University of Amsterdam, Science Park 904, 1098 XH Amsterdam, Netherlands}
\affil[i]{School of Physical Science and Technology, ShanghaiTech University, Shanghai 201210, China}
\affil[j]{Stanford Synchrotron Radiation Lightsource, SLAC National Accelerator Laboratory, Menlo Park, California 94025, USA}
\affil[k]{RIKEN SPring-8 Center, 1-1-1 Kouto, Sayo, Hyogo 679-5148, Japan}
\affil[l]{ICFO-Institut de Ci\`{e}ncies Fot\`{o}niques, The Barcelona Institute of Science and Technology, 08860
Castelldefels (Barcelona), Spain}
\affil[m]{Advanced Photon Source, Argonne National Laboratory, Argonne, Illinois 60439, USA}
\affil[n]{Max Planck Institute for the Structure and Dynamics of Matter, D-22761 Hamburg, Germany}
\affil[o]{Linac Coherent Light Source, SLAC National Accelerator Laboratory, Menlo Park, California 94025,
USA}
\affil[p]{RIKEN SPring-8 Center, 1-1-1 Kouto, Sayo, Hyogo 679-5148, Japan}
\affil[q]{Department of Physics and Astronomy, University of Tennessee, Knoxville, Tennessee 37996, USA}
\affil[r]{National Synchrotron Light Source II, Brookhaven National Laboratory, Upton, New York 11973, USA}
\affil[s]{Department of Physics and Astronomy, Aarhus University, Ny Munkegade 120, 8000 Aarhus C, Denmark}

\leadauthor{Mazzone, Meyer} 

\authorcontributions{M.P.M.D., X.L. and S.W.\ initiated and planned the project. D.G.M., D.M., Y.C. J.G.V., C.D.D., J.Q.L., V.T, Y.T., A.J., H.M., R.W., T.A.A., J.K, D.C., R.M., D.Z., R.A., S.S, H.Y., T.K, M.Y., Y.K., S.O., Y.T, J.L., J.Y., I.K.R., J.P.H., D.F.M., M.F., X.L., S.W., and M.P.M.D prepared for and performed the experiments, D.G.M., D.M., Y.C., S.W., and M.P.M.D. analyzed and interpreted the data. Y.S.\ and J.Y.\ synthesized the crystals and A.J.A.J., N.J.R., and R.M.K. provided theoretical support. D.G.M., D.M., X.L., S.W., and M.P.M.D.\ wrote the paper.}
\authordeclaration{The authors declare no competing interests.}

\equalauthors{\textsuperscript{1} D. G. M contributed equally to this work with D. M.}
\correspondingauthor{\textsuperscript{2}To whom correspondence should be addressed. E-mail: mdean@bnl.gov, liuxr@shanghaitec.edu.cn or simon.wall@icfo.eu.}

\significancestatement{Ultrafast manipulation of magnetic states holds great promise for progress in our understanding of new quantum states and technical applications, but our current knowledge of transient magnetism is very limited. Our work elucidates the nature of transient magnetism in gapped antiferromagnets using Sr$_3$Ir$_2$O$_7$ as a model material. We find that transient magnetic fluctuations are trapped throughout the entire Brillouin zone, while remaining present beyond the time that is required to restore the original spin network. The results are interpreted in the context of a spin-bottleneck effect, in which the existence of an explicit magnetic decay channel allows for an efficient thermalization of transient spin waves.}


\begin{abstract}
Although ultrafast manipulation of magnetism holds great promise for new physical phenomena and applications, targeting specific states is held back by our limited understanding of how magnetic correlations evolve on ultrafast timescales. Using ultrafast resonant inelastic x-ray scattering we demonstrate that femtosecond laser pulses can excite transient magnons at large wavevectors in gapped antiferromagnets, and that they persist for several picoseconds which is opposite to what is observed in nearly gapless magnets. Our work suggests that materials with isotropic magnetic interactions are preferred to achieve rapid manipulation of magnetism.
\end{abstract}

\dates{This manuscript was compiled on \today}
\doi{\url{www.pnas.org/cgi/doi/10.1073/pnas.XXXXXXXXXX}}

\begin{document}

\maketitle
\thispagestyle{firststyle}
\ifthenelse{\boolean{shortarticle}}{\ifthenelse{\boolean{singlecolumn}}{\abscontentformatted}{\abscontent}}{}

\section*{Introduction}
\begin{figure}[tbh]
\includegraphics[width=\linewidth]{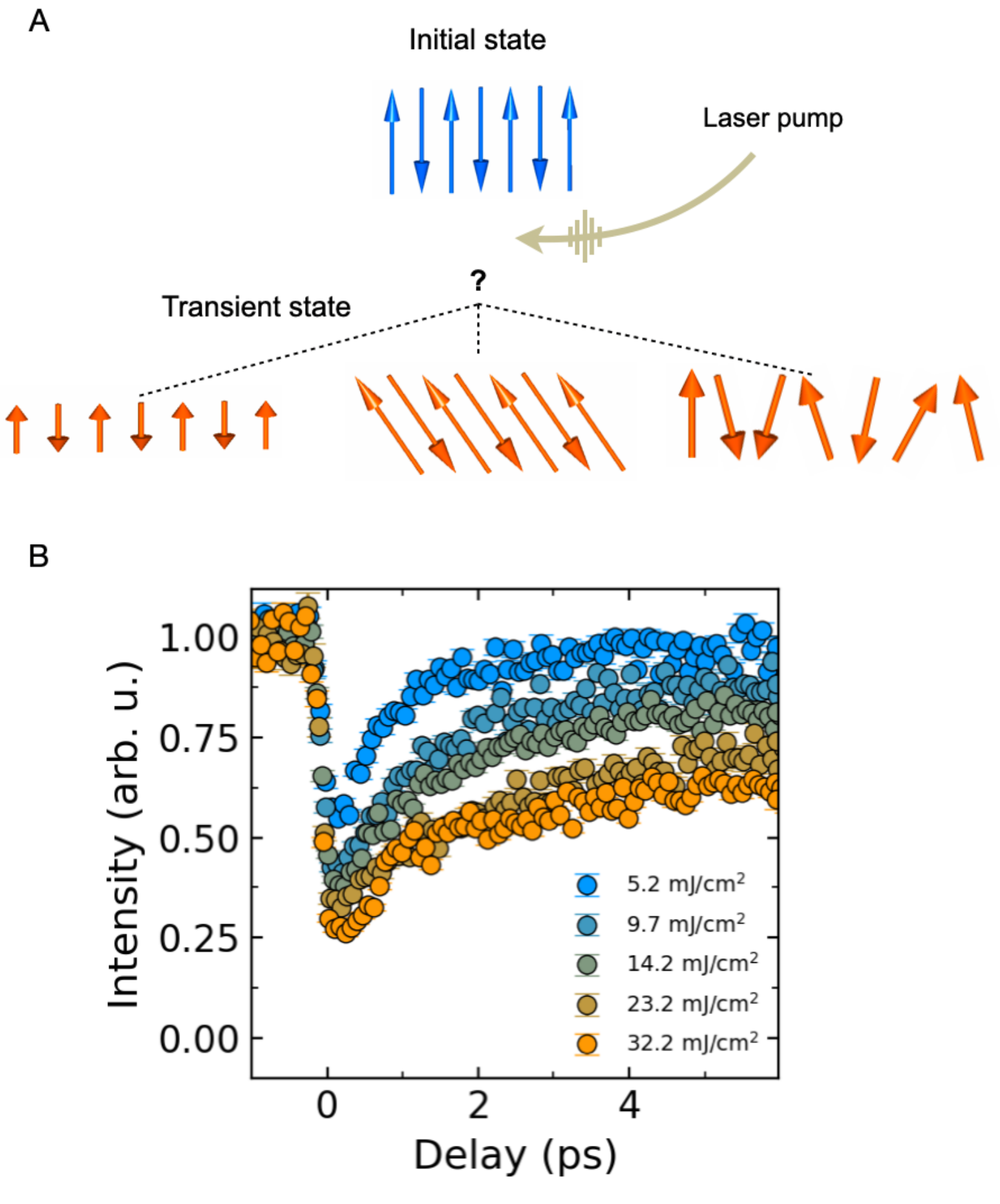}
\caption{Demagnetization pathways in antiferromagnets.  (A) Multiple spin configurations can give rise to the same macroscopic magnetization. These are indistinguishable in order parameter measurements commonly probed in ultrafast studies. Three such cases are sketched, corresponding to a reduced magnetic moment, a collective spin rotation and a disordered state. The first two cases are uniform perturbations to the entire spin network, whereas the latter is due to short range disorder of individual spins. (B) Relative magnetic (-3.5, 1.5, 18) Bragg peak intensity in Sr$_3$Ir$_2$O$_7$ as a function of time delay [notation in reciprocal lattice units (r.l.u)]. The data are plotted up to 7~ps after the arrival of the optical pump at $t$ = 0. The error bars follow Poissonian statistics. }
\label{REXS_data}
\end{figure}

\dropcap{U}ltrashort laser pulses are a powerful emerging tool to modify materials as they can induce new non-equilibrium states of matter \cite{Averitt2002, Dean2016, Basov2017, Gandolfi2017, Wang2018, Yue2017, Wall2018, Fausti2011light, Tzschaschel2019, Simoncig2017}. Particularly interesting is photo-excited magnetism, as in equilibrium, magnetic fluctuations are central to many phenomena including unconventional superconductivity, charge-stripe correlations and quantum spin liquids \cite{Keimer2017,Coslovich2013,Savary2016}. While increasing experimental evidence has shown that magnetism is suppressed by photo-doping, the exact nature of the spin configuration in the transient state and its evolution in time is largely unclear, holding back progress in our understanding. This is mainly because experimental tools for microscopically probing ultrafast magnetism are still in their infancy with respect to probing equilibrium magnetism. Most techniques are only sensitive to the magnetic order parameter or exclusively probe magnetic correlations at the Brillouin zone center. While often insightful, these probes cannot distinguish between various microscopic states, as their differences appear in the spacial (or Q) dependence of the magnetic correlations (see Fig.~\ref{REXS_data}A). This is important, for instance, in studies that focus on the transient behaviour of local exchange couplings, as they can be probed directly through magnons at the magnetic zone boundary. In fact, several theories predict that the magnetic exchange couplings among ordered ions change transiently under photo-excitation \cite{Secchi2013AnnalsPhysics,Bittner2018EDMFT,Mentinkrevew2017,Mentinknatcommun2015}. Thus, clarifying which magnons are excited in the transient state and how they evolve in time is crucial to unveil the detailed nature of transient states. 

We use time-resolved resonant x-ray scattering to overcome the aforementioned limitation, enabling studies of transient magnetic correlations throughout the Brillouin zone at ultrafast timescales. The technique uses incident x-rays that carry an appreciable momentum, and whose energy is tuned to a core-hole resonance, enhancing the sensitivity to magnetic modes. Magnetic long-range order is measured via \gls*{tr-REXS}. In contrast, \gls*{tr-RIXS} measures the energy loss and momentum change of scattered photons enabling to probe short-range magnetic fluctuations, such as magnons, in the transient state. \Gls*{tr-RIXS} measurements of magnons are extremely challenging technically. The first experiment was performed a few years previously on Sr$_2$IrO$_4$, but the only observable effects occurred at the magnetic ordering wavevector \cite{Dean2016}. This appears superficially congruent with the idea that the effectively zero momentum light can only excite individual magnons at the center of the magnetic Brillouin zone. Here we study gapped antiferromagnet Sr$_3$Ir$_2$O$_7$, and directly show that photo-excitation can modify magnons throughout the Brillouin zone which persist beyond a picosecond timescale.  We suggest that the large spin gap in Sr$_3$Ir$_2$O$_7$ blocks the cooling of transient magnons that appears to occur in Sr$_2$IrO$_4$.

\section*{Demagnetization pathways in antiferromagnets}

\begin{figure*}[tbh]
\includegraphics[width=\textwidth]{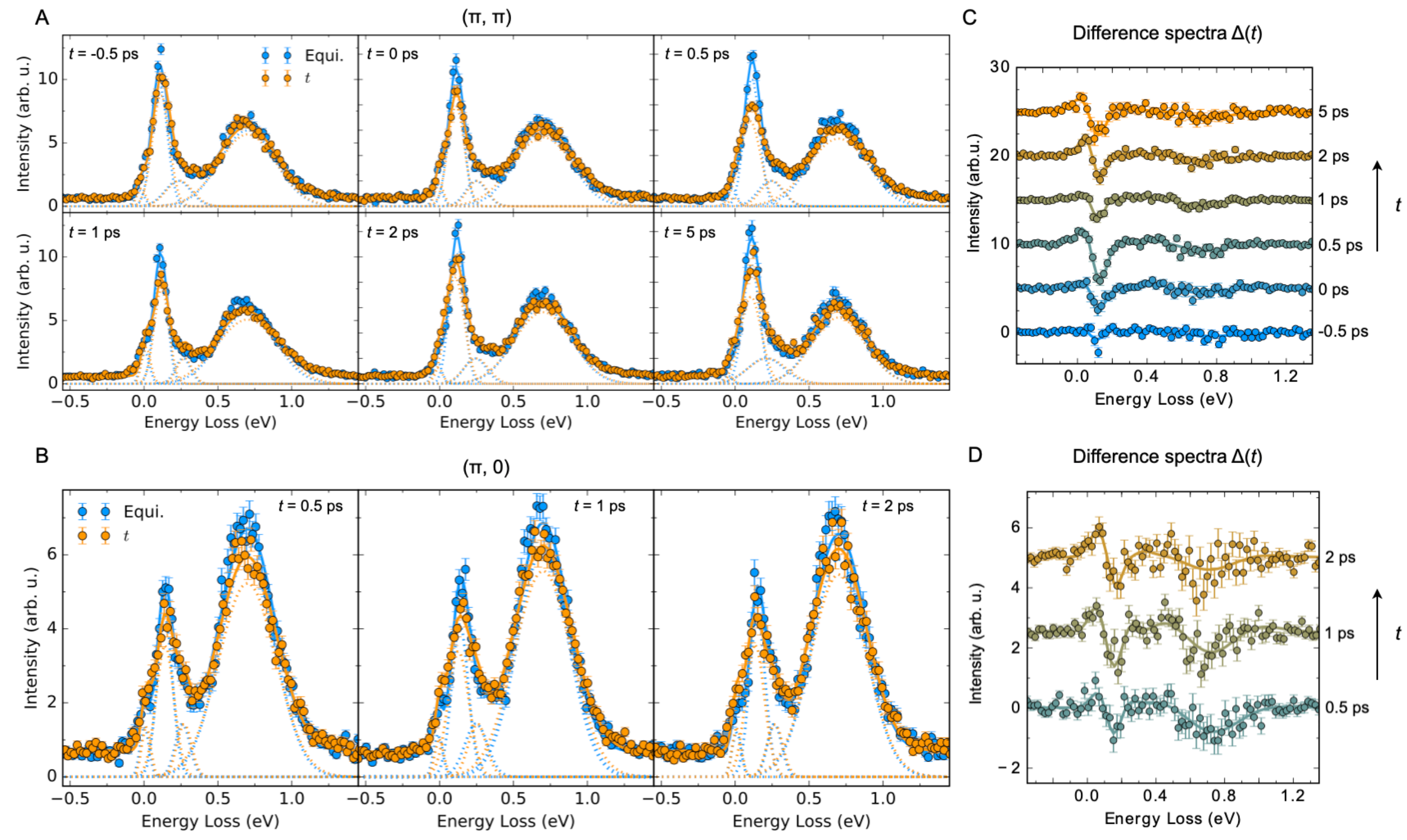} %
\caption{Static and transient electronic short-range correlations of Sr$_3$Ir$_2$O$_7$. Tr-RIXS spectra at the minimum (A) and maximum of the dispersion (B). The spectra show an elastic line, a magnon, a magnon continuum and an orbital excitation at $\sim$680~meV, before the arrival of the laser pump (blue) and for different time delays (orange). Errorbars are determined via Poissonian statistics. The dotted lines are contributions of best fits to the data (solid lines), consisting of four Gaussians and a constant background. Their difference spectra are shown in (C) and (D).
}
\label{RIXS_data}
\end{figure*}

Figure~\ref{REXS_data}B shows the time evolution of the Sr$_3$Ir$_2$O$_7$ magnetic Bragg peak intensity  up to 7~ps after the arrival of the 2~$\mu$m (620~meV) laser pump that excites carriers across the band-gap\cite{Moon2008}. This was measured using Ir $L_3$-ege \gls*{tr-REXS} at $T\approx 110$~K $\ll$ $T_N$ = 285~K using the (-3.5, 1.5, 18) magnetic Bragg peak (see Materials and Methods). Immediately after the pump, magnetism is reduced by 50-75\% for laser fluences between 5.2 and 32.2~mJ/cm$^2$. This occurs faster than our time resolution of 0.15~ps, which is substantially quicker than in other materials including Sr$_2$IrO$_4$ \cite{Afanasiev2019}. It is noted that even for strong fluences some remnant magnetic fraction is observed immediately after the laser pump, suggesting some experimental mismatch between pumped laser and probed x-ray volumes (c.f. SI Appendix). However, the transient signal shows signs of saturation, suggesting the pumped volume is completely demagnetized. We refer to further details on the recovery of the magnetic long-range order in the SI Appendix. The \gls*{tr-REXS} results in Fig.~\ref{REXS_data}B clearly show that a 2~$\mu$m laser pump results in an appreciable quenching of magnetic long-range order that survives on a picosecond timescale. However, the experiment does not reveal the microscopic spin configuration of the transient state. As illustrated in Fig.~\ref{REXS_data}A a transiently suppressed magnetic order parameter can arise from several very different microscopic configurations that give the same REXS response. We thus studied the RIXS response of the transient state to gain insight into the Q dependence of the magnetic correlations.
\section*{Transient evolution of magnetic short-range correlations}

Figures~\ref{RIXS_data}A and B display energy-loss spectra of Sr$_3$Ir$_2$O$_7$ measured at ($\pi,  \pi$) and $(\pi, 0)$, corresponding to the magnetic Brillouin zone center and zone boundary. The blue data points correspond to RIXS spectra in the equilibrium state, showing four features; an elastic line, a collective magnon, a magnon continuum and an orbital excitation \cite{Kim2012_2,Moretti2015,Lu2017,Lu2018}. The collective magnon features a moderate dispersion, shifting in energy from 100 meV at ($\pi,  \pi$) to 150 meV at $(\pi, 0)$, while the magnon continuum (250 meV) and orbital excitation (680 meV) reveal negligible dispersion. It is believed that the magnetic excitations in equilibrium arise predominantly from the combination of Heisenberg-like interactions within the iridium layers and an anisotropic exchange perpendicular to the tetragonal plane (see also SI Appendix) \cite{Cao2018,Kim2012,Kim2012_2}. The orbital excitation, on the other hand, can be understood considering the electronic configuration of Sr$_3$Ir$_2$O$_7$. It is defined by the five iridium valence electrons, residing in a crystal-field-derived $t_{2g}$ orbital manifold. The ground state is further split by strong spin-orbit coupling and moderate Coulomb interactions, establishing a filled $J_\text{eff}$ = 3/2 and half-filled $J_\text{eff}$ = 1/2 Mott state \cite{Cao2018,Kim2009,Kim2012,Kim2012_2}. Thus, the excitation epitomize a transition between these two manifolds \cite{Kim2012_2,Lu2017,Lu2018}.

We have overplotted the energy-loss spectra in equilibrium (blue) with transient state data (orange). These have been prepared with a 20~mJ/cm$^2$ laser pulse, for which a substantial suppression of magnetic order has been found (see Fig.~\ref{RIXS_data}B). Because small drifts in the X-ray source position can result into shifts of the measured RIXS spectrum, static and pumped spectra were measured with alternate shots of the free electron laser for each delay. The time resolution of the setup equaled $\sim$400 fs (see Materials and Methods for details). At both reciprocal lattice positions, we observe significant changes in the electronic correlations at $t> 0$. This is seen most clearly in the difference spectra shown in Figs.~\ref{RIXS_data}C and D, evincing that the magnon and orbital excitations are altered upon photo-doping at both, ($\pi,  \pi$) and $(\pi, 0)$. 

Further insight into the transient short-range correlations is gained from a quantitative analysis of the excitation spectra. A sum of four Gaussian-shaped peaks was used to represent the energy-loss data in Figs.~\ref{RIXS_data}A and B (see dotted lines in figure and the Materials and Methods section for further information). The fit shows that the magnon and orbital amplitudes are suppressed in the transient state and that their width is enlarged (see Fig.~\ref{RIXS_analysis}A and B). Notably, at the magnetic wavevector the fitting parameters are most strongly modified at 0.5~ps and recover only partially within 5~ps, which is in line with the incomplete recovery of magnetic long-range order (see Fig.~\ref{REXS_data}B).

\section*{Discussion}

The cross section of the orbital excitation around 680~meV is dependent on the respective charge occupations of the $J_\text{eff}$ = 3/2 and $J_\text{eff}$ = 1/2 state. The intensity would, for instance, decrease if fewer than four electrons per Ir side reside in the $J_\text{eff}$ = 3/2 ground state, and increase if the $J_\text{eff}$ = 1/2 state was empty. The transient depletion of the excitation thus provides direct evidence that not only $J_\text{eff}$ = 1/2 but also 3/2 electrons are pumped over the insulating band gap, and that some of the $J_\text{eff}$ = 3/2 carriers do not decay into their orbital ground state within 5~ps. This supports a picture in which the electronic subsystem remains in a transient state for appreciable time delays. 

\begin{figure*}[tbh]
\includegraphics[width=\textwidth]{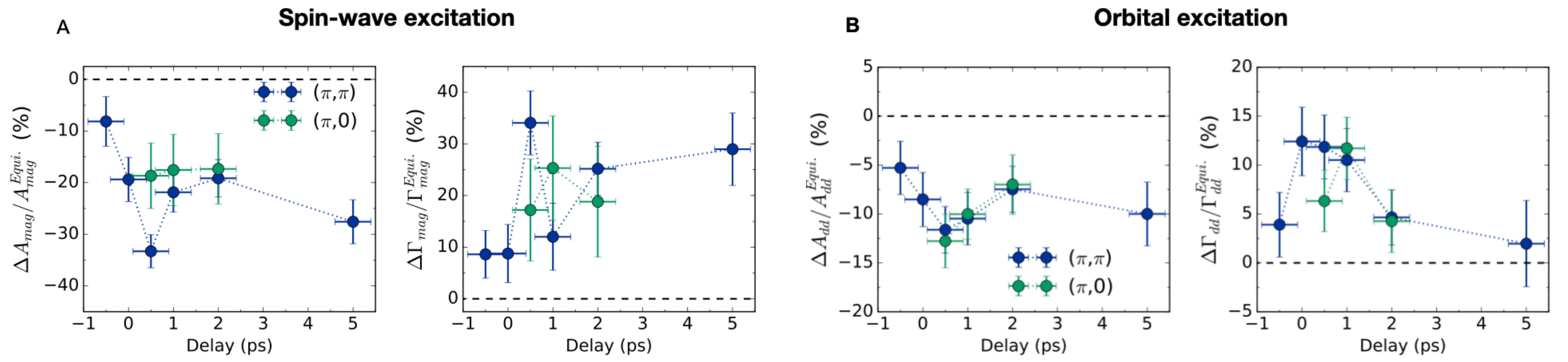} %
\caption{Ultrafast evolution of electronic correlations in Sr$_3$Ir$_2$O$_7$. Time dependence of the relative (A) magnon and (B) orbital amplitude, $A$, and full-width at half-maximum, $\Gamma$, respectively. The magnon and orbital amplitudes are suppressed in the transient state and their width is enlarged. Note the difference in scale. Errorbars are derived from the least-squares fitting algorithm.
}
\label{RIXS_analysis}
\end{figure*}

Having established the dynamics of the orbital excitation, we now turn to the transient behavior of the collective (pseudo-)spins. In equilibrium, magnetic long-range order ensures the existence of well-defined spin-waves. Upon photo-excitation, broader peaks are observed while the magnetic exchange couplings are unaffected. This can be conceptualized as a redistribution of magnons out of the modes populated in equilibrium, into a range of transient magnons around the original mode. Our data allow us to directly investigate how magnons are modified at different delays after photo-excitation, and to examine how they change across the Brillouin zone using ($\pi,  \pi$) and $(\pi, 0)$ as representative points.

\begin{figure}[tbh]
\includegraphics[width=\linewidth]{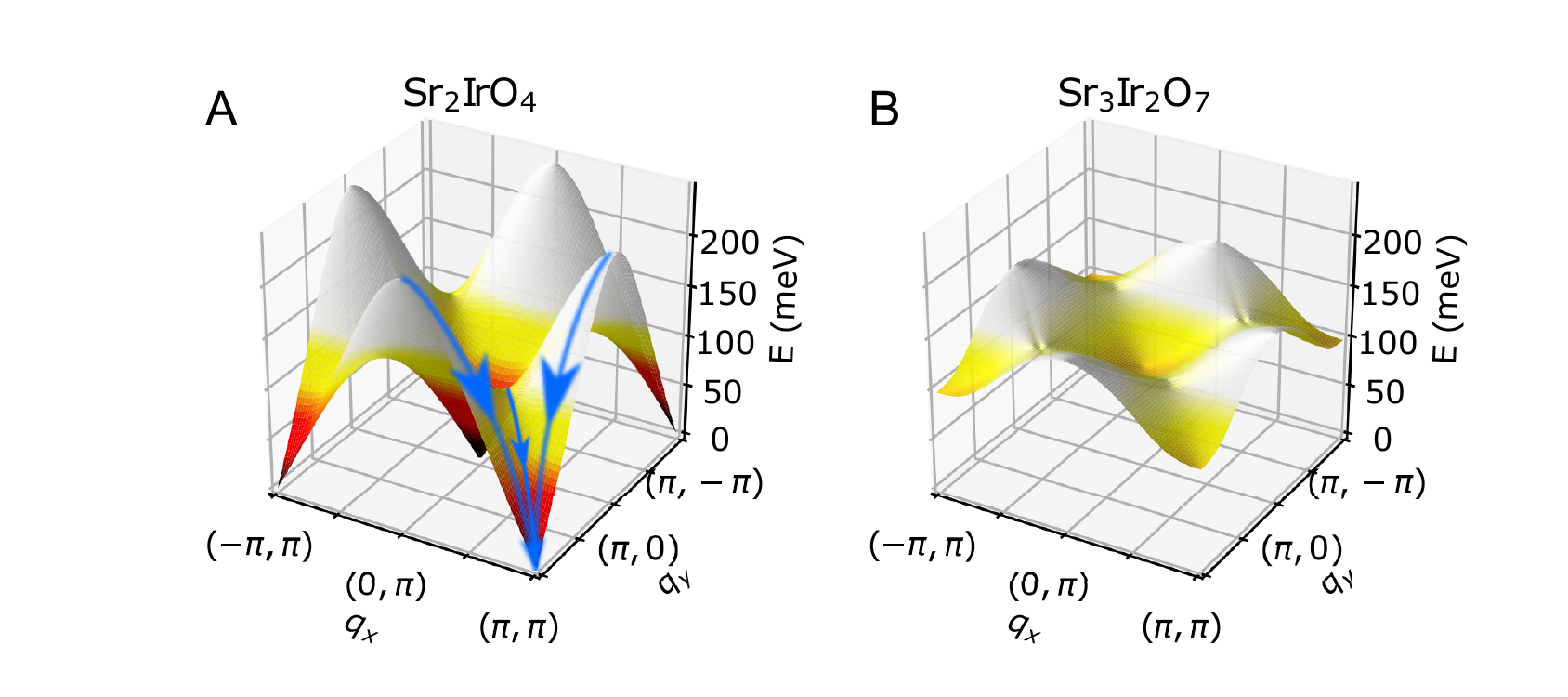}
\caption{Spin-bottleneck mechanism. (A) Heisenberg-like Sr$_2$IrO$_4$ features a near-zero-energy Goldstone mode establishing a well-defined channel (indicated schematically by the blue arrows) over which transient correlations decay into lower-energy multi-particle excitations. This leads to strong differences in the ultrafast magnetic response at ($\pi,  \pi$) and $(\pi, 0)$. (B) Such a decay channel is absent in gapped antiferromagnet Sr$_3$Ir$_2$O$_7$, leading to transient magnons that are trapped in the entire Brillouin zone.}
\label{dispersion}
\end{figure}

The substantial anisotropy perpendicular to the tetragonal plane leads to collinear magnetic long-range order, revealing a large magnon gap that is minimal at ($\pi,  \pi$) and maximal at $(\pi, 0)$ (cf. Fig.~\ref{dispersion}B for details) \cite{Cao2018,Kim2012_2,Moretti2015,Lu2017,Lu2018}. Thus, the results in Fig.~\ref{RIXS_analysis}A show that magnons are modified over an extended area in reciprocal space, and that these changes persist for several picoseconds. This experimental result is strikingly different from the behavior in materials hosting a gapless excitation spectrum. In isotropic Heisenberg-like Sr$_2$IrO$_4$ (see Fig.~\ref{dispersion}A), for instance, the only observable changes in the magnon occurred at the magnetic Brillouin zone center of $(\pi, \pi)$ \cite{Dean2016}. In the following discussion, we aim to assess these differences.

The transient state created here features suppressed magnetic long- and short-range correlations throughout the Brillouin zone and it, therefore, has some similarities to a thermal state at elevated temperature. This state, however, appears to form effectively instantaneously, within our time resolution of either 150 or 400~fs depending on whether one considers the REXS or RIXS signatures. These timescales are much faster than typical thermalization processes in magnetic insulators, ranging from several to hundreds of picoseconds  \cite{Kimel2002, Kimel2006, Matsubara2007, Yada2011}. We thus consider the physical mechanisms at play, beyond a simple effective temperature model.

The photo-excitation process promoting charge carriers across the insulating gap leaves behind local spin vacancies that create non-thermal magnons spread across the entire Brillouin zone. These high-energy magnons can decay into lower-energy multiple-particle excitations via magnon-magnon scattering processes following the dispersion relation - which, in lowest approximation, is similar to the one of the unperturbed state. A full thermalization of the system, however, requires a coupling other subsystems, which becomes unavoidable for the dissipation of transient magnons around the minimum of the dispersion relation. 

In gapless antiferromagnets, such as Sr$_2$IrO$_4$ \cite{Calder2018,Kim2012}, a steep magnon dispersion from high to effectively zero energy establishes a well-defined decay channel (see Fig.~\ref{dispersion}a). Thus, spin waves at high energy face multiple possibilities to disintegrate, but the options become increasingly limited around the minimum of the Goldstone-like mode. This gives a natural explanation of why ultrafast magnons were observed only at ($\pi,  \pi$), but not at $(\pi, 0)$ for Sr$_2$IrO$_4$ \cite{Dean2016}. Furthermore, the small spin-wave gap enables an efficient energy transfer into the lattice subsystem via low-energy phonons, restoring the equilibrium configuration. 

We note that N\'{e}el order in Sr$_2$IrO$_4$ is unstable if one considers only a single Ir-layer, but stabilized by a very small inter-layer c-axis coupling amplified by a divergence in the in-plane magnetic correlation length within the Ir-layers (see SI Appendix). This is in strong contrast to bi-layer materials such as Sr$_3$Ir$_2$O$_7$ which order magnetically without necessarily invoking coupling between different bi-layers. Although the bi-layer system features far stronger anisotropy and intra-bilayer coupling, this is offset by the spin gap impeding the growth of the in-plane magnetic correlation proceeding the transition. Empirically, Sr$_3$Ir$_2$O$_7$ and Sr$_2$IrO$_4$ feature comparable N\'{e}el temperatures (285 and 240~K, respectively) despite the spin-wave gap of the former material being roughly 150 times larger (see Fig.~\ref{dispersion}) \cite{Calder2018,Kim2012,Kim2012_2}. Thus, no pronounced decay channel is present in Sr$_3$Ir$_2$O$_7$, which inherently limits the decay of transient magnons along a particular wavevector. Furthermore, phonon-assisted energy transfer processes are strongly reduced by the large spin gap, leading to long-lived magnons that are trapped over a large reciprocal space area.

Following this argument, the partial recovery of magnetic long-range order at a picosecond timescale may arise from two mechanisms. The magnetic time constant is close to the charge timescale that has been reported previously via ultrafast reflectivity measurements \cite{Chu2017}. This may attest a link between charge and magnetic degrees of freedom, which is also supported by the wavevector-independent suppression of the orbital excitation below 2~ps (see Fig.~\ref{RIXS_analysis}B). In Mott materials the (pseudo-)spin configuration is directly related to the charge distribution on the atomic sites, whose recovery reinstates the original one spin/site population. Thus, it is conceivable to attribute the partial recovery of magnetic order in Sr$_3$Ir$_2$O$_7$ to its Mott nature. Alternatively, the partial recovery of magnetic long-range order may originate from a redistribution of transient magnons within the Brillouin zone, pointing towards a scenario in which the laser pump generates an excess of magnons at the magnetic wavevector. In fact, the results shown in Fig.~\ref{RIXS_analysis}A suggest a suppression of spin waves at ($\pi,  \pi$), which is absent at $(\pi, 0)$.  

In summary, we report a near instantaneous creation of transient magnons in the gapped antiferromagnet Sr$_3$Ir$_2$O$_7$ after a laser excitation of  2~$\mu$m. The ultrafast transverse short-range correlations occur at both extremes of the magnetic dispersion in momentum space, showing the existence of trapped spin-waves that are likely present throughout the entire Brillouin zone. This demonstrates an incoherent response of transient magnetism, which is fundamentally different than in gapless antiferromagnets such as in Heisenberg-like Sr$_2$IrO$_4$ \cite{Dean2016}. The results are interpreted in the context of a spin-bottleneck effect. In Sr$_2$IrO$_4$, a steep dispersion exists from high energy magnons at the magnetic zone boundary into low-energy magnons at $(\pi, \pi)$, which allows for a highly efficient magnon decay. In contrast, the large magnon gap and moderate dispersion of Sr$_3$Ir$_2$O$_7$ leads to transient spin waves that are trapped in the entire Brillouin zone. Thus, our results emphasize the need to include contributions like anisotropic magnetic exchange interactions alongside electron-phonon couplings in theoretical approaches that model transient magnetic states. Finally, we remind the reader that neither a collinear antiferromagnetic spin alignment, nor magnons are exact eigenstates of a $S = 1/2$ quantum magnet. Thus, it may be natural to expect strong magnon-magnon scattering occurs in some transient states. Further experimental and theoretical efforts along these lines will be vital to progress in the understanding of the ultrafast dynamics in quantum materials, including cases where complex competitions of macroscopic quantum phases appear via photo-doping.
\matmethods{
\textbf{Sample preparation and characterization.} All measurements were performed on Sr$_3$Ir$_2$O$_7$ single crystals that were grown with a flux method as described in Ref.~\cite{Li2013}, and references therein. The high quality of the crystals were confirmed via Laue and single crystal diffraction. Here, we denote reciprocal space using a pseudo-tetragonal unit cell with $a$ = 3.896~\AA~and $c$ = 20.88~\AA~at room temperature.

All experiments were conducted on the same sample with crystal orientations $[1, 0, 0]$ and $[0, 0, 1]$, or $[1, 1, 0]$ and $[0, 0, 1]$ in the scattering plane for \gls*{tr-REXS} or \gls*{tr-RIXS} experiments, respectively. In both cases, a horizontal scattering geometry was used and the sample was cooled to $T\approx$ 110~K via nitrogen cryostreams.

\textbf{Optical laser pump.} 100~fs long pump pulses of 620~meV (2~$\mu$m) were generated by an optical parametric amplifier similar to previous ultrafast experiments on Sr$_2$IrO$_4$ \cite{Dean2016}. The angle between the incident x-ray and infrared pulses was 20$^\circ$. This leads to an estimated laser penetration depth of $\sim82$~nm for grazing incident x-rays (see SI Appendix).

\textbf{Time-resolved resonant elastic scattering (tr-REXS).} The ultrafast recovery of magnetic long-range order was studied at the beamline~3 of the SPring-8 Angstrom Compact free-electron LAser (SACLA) featuring a repetition rate of 30~Hz \cite{Huang2012,Tono2013}. The x-ray energy was tuned to the Ir $L_3$-edge at 11.215~keV, and refined via a resonant energy scan of the $(-3.5, 1.5, 18)$ magnetic Bragg peak. The reflection was detected at 2$\theta$ = 92.2$^\circ$ using a multi-port charged coupled device (MPCCD) area detector (0.0086~r.l.u/pixel) \cite{Kameshima2014}. The Bragg condition was established by rotating the sample surface normal $\Delta\chi$ = 21.2$^\circ$ out of the scattering plane. The x-rays were focused to a spot size of 20~$\mu$m full-width at half-maximum (FWHM) and hit the sample at a grazing angle of $\sim$1.6$^\circ$.The x-ray penetration depth was estimated to $162$~nm (see SI Appendix). The laser spot size was $\sim$230~$\mu$m FWHM at normal incidence. The detector was read out shot-by-shot and thresholded to reduce background arising from x-ray fluoresence and electrical noise. Two-dimensional detector images were integrated around the position where the Bragg peak was observed. Each data point contains statistics from 300-600 shots. The 300~fs jitter of the free electron laser was corrected with a GaAs timing tool for delays $t$ $<$ 7~ps, yielding an effective time resolution of 150~fs at FWHM \cite{Katayama2016}.

Figure \ref{REXS_data}B reports the magnetic Bragg peak ratio $I_\text{on}$/$I_\text{off}$, where $I_\text{on}$ denotes the intensity after laser excitation and $I_\text{off}$ when the laser was switched off. It is noted that even for strong fluences some remnant magnetic signal is observed for $t$ $>$ 0, suggesting some experimental mismatch between pumped laser and probed x-ray volumes Further information on the time evolution of the magnetic order parameter is given in the SI Appendix.  

\textbf{Time-resolved resonant inelastic scattering (tr-RIXS).} The ultrafast recovery of magnetic short-range correlations was studied at the X-ray Pump Probe (XPP) instrument of the Linac Coherent Light Source (LCLS) running at a repetition rate of 120~Hz \cite{Chollet2015}. The data at ($\pi,  \pi$) and $(\pi, 0)$ correspond to measurements at $(-3.5, 0.5, 20)$ and $(-3.5, 0, 19.6)$ with $2\theta = 93.7$ and $91.8^\circ$. The x-rays were focused to a spot size of 50~$\mu$m FWHM. We used a grazing incident geometry of 3.7 and 2.1$^\circ$ for the two reciprocal lattice positions, yielding an  x-ray penetration depth of $378$ and $214$~nm, respectively (see Supplemental Information). The laser spot size equaled 845 $\mu$m FWHM at normal incidence and the fluence was fixed to 20~mJ/cm$^2$. The same RIXS spectrometer as in Ref.~\cite{Dean2016} was used, delivering an energy resolution of $\sim70$~meV with an angular acceptance of $6^\circ$ (0.28 \AA$^{-1}$) on the analyser. The  ePix100a area detector was read out every shot \cite{Carini2016}. The time-resolution of the experiment arising from a combined jitter and beam drift effect was limited to $\sim$0.4~ps. It is noted that it proved impractical to use the timing tool, as it was insufficiently sensitive on the aggressively monochromatized incident beam. RIXS spectra were collected in a stationary mode of 144000 shots each. Two to six acquisitions were taken for each time delay. The data presented in Fig.~\ref{RIXS_data}A and B were normalized to an incident beam intensity monitor and calibrated to the energy of the orbital excitation at 680~meV \cite{Kim2012_2,Moretti2015,Lu2017,Lu2018}. The center of the elastic line and its width were fixed in the analysis to 0~meV and the experimental resolution, respectively. Robust fits were obtained by further constraining the energy of the magnon continuum to 260~meV and its amplitude to the mean values of 2.02 and 1.44 for ($\pi,  \pi$) and $(\pi, 0)$, respectively. All other parameters were varied freely, and shown in the SI appendix, Fig. S2.
}

\showmatmethods{} 

\acknow{We thank Chris Homes for discussions. The design, execution of the experiments, as well as the data analysis are supported by the U.S.\ Department of Energy, Office of Science, Basic Energy Sciences, Materials Science and Engineering Division, under Contract No.\ DE-SC0012704(BNL) and DE-AC02-06CH11357(ANL). D.G.M.\ acknowledges funding from the Swiss National Science Foundation, Fellowship No.\ P2EZP2\_175092. C.D.D.\ was supported by the Engineering and Physical Sciences Research Council (EPSRC) Centre for Doctoral Training in the Advanced Characterisation of Materials under Grant No.\ EP/L015277/1. X. L.\ and J.Q.L.\ were supported by the ShanghaiTech University startup fund, MOST of China under Grant No.\ 2016YFA0401000, NSFC under Grant No.\ 11934017 and the Chinese Academy of Sciences under Grant No.\ 112111KYSB20170059. Work at UCL was supported by the EPSRC under Grants No. EP/N027671/1 and No. EP/N034872/1. The work at ICFO received financial support from the Spanish Ministry of Economy and Competitiveness through the ``Severo Ochoa'' program for Centres of Excellence in R\&D (SEV-2015-0522), from Fundaci\'{o} Privada Cellex, from Fundaci\'{o} Mir-Puig, and from Generalitat de Catalunya through the CERCA program and from the European Research Council (ERC) under the European Union’s Horizon 2020 research and innovation programme (Grant agreement No.~758461). J.L.\ acknowledges support from the National Science Foundation under Grant No.~DMR-1848269. The magnetic Bragg peak measurements were performed at the BL3 of SACLA with the approval of the Japan Synchrotron Radiation Research Institute (JASRI) (Proposal No.\ 2017A8077, 2018A8032 and 2019A8087). This research made use of the Linac Coherent Light Source (LCLS), SLAC National Accelerator Laboratory, which is a DOE Office of Science User Facility, under Contract No.\ DE-AC02-76SF00515.}

\showacknow{} 

\bibliography{refs}

\begin{thebibliography}{39}
\providecommand{\natexlab}[1]{#1}
\providecommand{\url}[1]{\texttt{#1}}
\expandafter\ifx\csname urlstyle\endcsname\relax
  \providecommand{\doi}[1]{doi: #1}\else
  \providecommand{\doi}{doi: \begingroup \urlstyle{rm}\Url}\fi

\bibitem[Averitt and Taylor(2002)]{Averitt2002}
R~D Averitt and A~J Taylor.
\newblock Ultrafast optical and far-infrared quasiparticle dynamics in
  correlated electron materials.
\newblock \emph{Journal of Physics: Condensed Matter}, 14\penalty0
  (50):\penalty0 R1357--R1390, dec 2002.
\newblock \doi{10.1088/0953-8984/14/50/203}.
\newblock URL \url{https://doi.org/10.1088%2F0953-8984%2F14%2F50%2F203}.

\bibitem[Dean et~al.(2016)Dean, Cao, Liu, Wall, Zhu, Mankowsky, Thampy, Chen,
  Vale, Casa, Kim, Said, Juhas, Alonso-Mori, Glownia, Robert, Robinson,
  Sikorski, Song, Kozina, Lemke, Patthey, Owada, Katayama, Yabashi, Tanaka,
  Togashi, Liu, Rayan~Serrao, Kim, Huber, Chang, McMorrow, F{\"o}rst, and
  Hill]{Dean2016}
M.~P.~M. Dean, Y.~Cao, X.~Liu, S.~Wall, D.~Zhu, R.~Mankowsky, V.~Thampy, X.~M.
  Chen, J.~G. Vale, D.~Casa, Jungho Kim, A.~H. Said, P.~Juhas, R.~Alonso-Mori,
  J.~M. Glownia, A.~Robert, J.~Robinson, M.~Sikorski, S.~Song, M.~Kozina,
  H.~Lemke, L.~Patthey, S.~Owada, T.~Katayama, M.~Yabashi, Yoshikazu Tanaka,
  T.~Togashi, J.~Liu, C.~Rayan~Serrao, B.~J. Kim, L.~Huber, C.~L. Chang, D.~F.
  McMorrow, M.~F{\"o}rst, and J.~P. Hill.
\newblock Ultrafast energy- and momentum-resolved dynamics of magnetic
  correlations in the photo-doped mott insulator {Sr$_2$IrO$_4$}.
\newblock \emph{Nature Materials}, 15:\penalty0 601--605, 05 2016.

\bibitem[Basov et~al.(2017)Basov, Averitt, and Hsieh]{Basov2017}
D.~N. Basov, R.~D. Averitt, and D.~Hsieh.
\newblock Towards properties on demand in quantum materials.
\newblock \emph{Nature Materials}, 16:\penalty0 1077 EP --, 10 2017.

\bibitem[Gandolfi et~al.(2017)Gandolfi, Celardo, Borgonovi, Ferrini, Avella,
  Banfi, and Giannetti]{Gandolfi2017}
M~Gandolfi, G~L Celardo, F~Borgonovi, G~Ferrini, A~Avella, F~Banfi, and
  C~Giannetti.
\newblock Emergent ultrafast phenomena in correlated oxides and
  heterostructures.
\newblock \emph{Physica Scripta}, 92\penalty0 (3):\penalty0 034004, jan 2017.

\bibitem[Wang et~al.(2018)Wang, Claassen, Pemmaraju, Jia, Moritz, and
  Devereaux]{Wang2018}
Yao Wang, Martin Claassen, Chaitanya~Das Pemmaraju, Chunjing Jia, Brian Moritz,
  and Thomas~P. Devereaux.
\newblock Theoretical understanding of photon spectroscopies in correlated
  materials in and out of equilibrium.
\newblock \emph{Nature Reviews Materials}, 3\penalty0 (9):\penalty0 312--323,
  2018.

\bibitem[Cao et~al.(2019)Cao, Mazzone, Meyers, Hill, Liu, Wall, and
  Dean]{Yue2017}
Y.~Cao, D.~G. Mazzone, D.~Meyers, J.~P. Hill, X.~Liu, S.~Wall, and M.~P.~M.
  Dean.
\newblock Ultrafast dynamics of spin and orbital correlations in quantum
  materials: an energy- and momentum-resolved perspective.
\newblock \emph{Philosophical Transactions of the Royal Society A:
  Mathematical, Physical and Engineering Sciences}, 377\penalty0
  (2145):\penalty0 20170480, 2019.

\bibitem[Wall et~al.(2018)Wall, Yang, Vidas, Chollet, Glownia, Kozina,
  Katayama, Henighan, Jiang, Miller, Reis, Boatner, Delaire, and
  Trigo]{Wall2018}
Simon Wall, Shan Yang, Luciana Vidas, Matthieu Chollet, James~M. Glownia,
  Michael Kozina, Tetsuo Katayama, Thomas Henighan, Mason Jiang, Timothy~A.
  Miller, David~A. Reis, Lynn~A. Boatner, Olivier Delaire, and Mariano Trigo.
\newblock Ultrafast disordering of vanadium dimers in photoexcited {VO$_2$}.
\newblock \emph{Science}, 362\penalty0 (6414):\penalty0 572--576, 2018.

\bibitem[Fausti et~al.(2011)Fausti, Tobey, Dean, Kaiser, Dienst, Hoffmann,
  Pyon, Takayama, Takagi, and Cavalleri]{Fausti2011light}
D.~Fausti, R.~I. Tobey, N.~Dean, S.~Kaiser, A.~Dienst, M.~C. Hoffmann, S.~Pyon,
  T.~Takayama, H.~Takagi, and A.~Cavalleri.
\newblock Light-induced superconductivity in a stripe-ordered cuprate.
\newblock \emph{Science}, 331\penalty0 (6014):\penalty0 189--191, 2011.
\newblock ISSN 0036-8075.
\newblock \doi{10.1126/science.1197294}.
\newblock URL \url{https://science.sciencemag.org/content/331/6014/189}.

\bibitem[Tzschaschel et~al.(2019)Tzschaschel, Satoh, and
  Fiebig]{Tzschaschel2019}
Christian Tzschaschel, Takuya Satoh, and Manfred Fiebig.
\newblock Tracking the ultrafast motion of an antiferromagnetic order
  parameter.
\newblock \emph{Nature Communications}, 10\penalty0 (1):\penalty0 3995, 2019.
\newblock \doi{10.1038/s41467-019-11961-9}.
\newblock URL \url{https://doi.org/10.1038/s41467-019-11961-9}.

\bibitem[Simoncig et~al.(2017)Simoncig, Mincigrucci, Principi, Bencivenga,
  Calvi, Foglia, Kurdi, Matruglio, Dal~Zilio, Masciotti, Lazzarino, and
  Masciovecchio]{Simoncig2017}
A.~Simoncig, R.~Mincigrucci, E.~Principi, F.~Bencivenga, A.~Calvi, L.~Foglia,
  G.~Kurdi, A.~Matruglio, S.~Dal~Zilio, V.~Masciotti, M.~Lazzarino, and
  C.~Masciovecchio.
\newblock Generation of coherent magnons in nio stimulated by euv pulses from a
  seeded free-electron laser.
\newblock \emph{Phys. Rev. Materials}, 1:\penalty0 073802, Dec 2017.
\newblock \doi{10.1103/PhysRevMaterials.1.073802}.
\newblock URL \url{https://link.aps.org/doi/10.1103/PhysRevMaterials.1.073802}.

\bibitem[Keimer and Moore(2017)]{Keimer2017}
B.~Keimer and J.~E. Moore.
\newblock The physics of quantum materials.
\newblock \emph{Nature Physics}, 13:\penalty0 1045 EP --, 10 2017.

\bibitem[Coslovich et~al.(2013)Coslovich, Huber, Lee, Chuang, Zhu, Sasagawa,
  Hussain, Bechtel, Martin, Shen, Schoenlein, and Kaindl]{Coslovich2013}
G.~Coslovich, B.~Huber, W.~S. Lee, Y.~D. Chuang, Y.~Zhu, T.~Sasagawa,
  Z.~Hussain, H.~A. Bechtel, M.~C. Martin, Z.~X. Shen, R.~W. Schoenlein, and
  R.~A. Kaindl.
\newblock Ultrafast charge localization in a stripe-phase nickelate.
\newblock \emph{Nature Communications}, 4:\penalty0 2643 EP --, 10 2013.

\bibitem[Savary and Balents(2016)]{Savary2016}
Lucile Savary and Leon Balents.
\newblock Quantum spin liquids: a review.
\newblock \emph{Reports on Progress in Physics}, 80\penalty0 (1):\penalty0
  016502, nov 2016.

\bibitem[Secchi et~al.(2013)Secchi, Brener, Lichtenstein, and
  Katsnelson]{Secchi2013AnnalsPhysics}
A.~Secchi, S.~Brener, A.I. Lichtenstein, and M.I. Katsnelson.
\newblock Non-equilibrium magnetic interactions in strongly correlated systems.
\newblock \emph{Annals of Physics}, 333:\penalty0 221, 2013.

\bibitem[Bittner et~al.(2018)Bittner, Gole\v{z}, Strand, Eckstein, and
  Werner]{Bittner2018EDMFT}
Nikolaj Bittner, Denis Gole\v{z}, Hugo U.~R. Strand, M.~Eckstein, and
  P.~Werner.
\newblock Coupled charge and spin dynamics in a photo-excited mott insulator.
\newblock \emph{Phys. Rev. B}, 97:\penalty0 235125, 2018.

\bibitem[Mentink(2017)]{Mentinkrevew2017}
J.~H. Mentink.
\newblock Manipulating magnetism by ultrafast control of the exchange
  interaction.
\newblock \emph{J. Phys.: Condens. Matter}, 23:\penalty0 453001, 2017.

\bibitem[Mentink et~al.(2015)Mentink, Balzer, and
  Echkstein]{Mentinknatcommun2015}
J.~H. Mentink, K.~Balzer, and M.~Echkstein.
\newblock Ultrafast and reversible control of the exchange interaction in mott
  insulators.
\newblock \emph{Nat. Commun.}, 6:\penalty0 6708, 2015.

\bibitem[Moon et~al.(2008)Moon, Jin, Kim, Choi, Lee, Yu, Cao, Sumi, Funakubo,
  Bernhard, and Noh]{Moon2008}
S.~J. Moon, H.~Jin, K.~W. Kim, W.~S. Choi, Y.~S. Lee, J.~Yu, G.~Cao, A.~Sumi,
  H.~Funakubo, C.~Bernhard, and T.~W. Noh.
\newblock Dimensionality-controlled insulator-metal transition and correlated
  metallic state in $5d$ transition metal oxides {Sr$_{n+1}$Ir$_n$O$_{3n+1}$}:
  ($n=1$, 2, and $\ensuremath{\infty}$).
\newblock \emph{Phys. Rev. Lett.}, 101:\penalty0 226402, Nov 2008.

\bibitem[Afanasiev et~al.(2019)Afanasiev, Gatilova, Groenendijk, Ivanov,
  Gibert, Gariglio, Mentink, Li, Dasari, Eckstein, Rasing, Caviglia, and
  Kimel]{Afanasiev2019}
D.~Afanasiev, A.~Gatilova, D.~J. Groenendijk, B.~A. Ivanov, M.~Gibert,
  S.~Gariglio, J.~Mentink, J.~Li, N.~Dasari, M.~Eckstein, Th. Rasing, A.~D.
  Caviglia, and A.~V. Kimel.
\newblock Ultrafast spin dynamics in photodoped spin-orbit mott insulator
  {Sr$_2$IrO$_4$}.
\newblock \emph{Phys. Rev. X}, 9:\penalty0 021020, Apr 2019.

\bibitem[Kim et~al.(2012{\natexlab{a}})Kim, Said, Casa, Upton, Gog, Daghofer,
  Jackeli, van~den Brink, Khaliullin, and Kim]{Kim2012_2}
Jungho Kim, A.~H. Said, D.~Casa, M.~H. Upton, T.~Gog, M.~Daghofer, G.~Jackeli,
  J.~van~den Brink, G.~Khaliullin, and B.~J. Kim.
\newblock Large spin-wave energy gap in the bilayer iridate
  {Sr$_3$Ir$_2$O$_7$}: Evidence for enhanced dipolar interactions near the
  {Mott} metal-insulator transition.
\newblock \emph{Phys. Rev. Lett.}, 109:\penalty0 157402, Oct
  2012{\natexlab{a}}.

\bibitem[Moretti~Sala et~al.(2015)Moretti~Sala, Schnells, Boseggia, Simonelli,
  Al-Zein, Vale, Paolasini, Hunter, Perry, Prabhakaran, Boothroyd, Krisch,
  Monaco, R\o{}nnow, McMorrow, and Mila]{Moretti2015}
M.~Moretti~Sala, V.~Schnells, S.~Boseggia, L.~Simonelli, A.~Al-Zein, J.~G.
  Vale, L.~Paolasini, E.~C. Hunter, R.~S. Perry, D.~Prabhakaran, A.~T.
  Boothroyd, M.~Krisch, G.~Monaco, H.~M. R\o{}nnow, D.~F. McMorrow, and
  F.~Mila.
\newblock Evidence of quantum dimer excitations in {Sr$_3$Ir$_2$O$_7$}.
\newblock \emph{Phys. Rev. B}, 92:\penalty0 024405, Jul 2015.

\bibitem[Lu et~al.(2017)Lu, McNally, Moretti~Sala, Terzic, Upton, Casa, Ingold,
  Cao, and Schmitt]{Lu2017}
Xingye Lu, D.~E. McNally, M.~Moretti~Sala, J.~Terzic, M.~H. Upton, D.~Casa,
  G.~Ingold, G.~Cao, and T.~Schmitt.
\newblock Doping evolution of magnetic order and magnetic excitations in
  {(Sr$_{1-x}$La$_x$)$_3$Ir$_2$O$_7$}.
\newblock \emph{Phys. Rev. Lett.}, 118:\penalty0 027202, Jan 2017.

\bibitem[Lu et~al.(2018)Lu, Olalde-Velasco, Huang, Bisogni, Pelliciari, Fatale,
  Dantz, Vale, Hunter, Chang, Strocov, Perry, Grioni, McMorrow, R\o{}nnow, and
  Schmitt]{Lu2018}
Xingye Lu, Paul Olalde-Velasco, Yaobo Huang, Valentina Bisogni, Jonathan
  Pelliciari, Sara Fatale, Marcus Dantz, James~G. Vale, E.~C. Hunter, Johan
  Chang, Vladimir~N. Strocov, R.~S. Perry, Marco Grioni, D.~F. McMorrow,
  Henrik~M. R\o{}nnow, and Thorsten Schmitt.
\newblock Dispersive magnetic and electronic excitations in iridate perovskites
  probed by oxygen {$K$}-edge resonant inelastic x-ray scattering.
\newblock \emph{Phys. Rev. B}, 97:\penalty0 041102, Jan 2018.

\bibitem[Cao and Schlottmann(2018)]{Cao2018}
Gang Cao and Pedro Schlottmann.
\newblock The challenge of spin{\textendash}orbit-tuned ground states in
  iridates: a key issues review.
\newblock \emph{Reports on Progress in Physics}, 81\penalty0 (4):\penalty0
  042502, feb 2018.

\bibitem[Kim et~al.(2012{\natexlab{b}})Kim, Casa, Upton, Gog, Kim, Mitchell,
  van Veenendaal, Daghofer, van~den Brink, Khaliullin, and Kim]{Kim2012}
Jungho Kim, D.~Casa, M.~H. Upton, T.~Gog, Young-June Kim, J.~F. Mitchell,
  M.~van Veenendaal, M.~Daghofer, J.~van~den Brink, G.~Khaliullin, and B.~J.
  Kim.
\newblock Magnetic excitation spectra of {Sr$_2$IrO$_4$}: Probed by resonant
  inelastic x-ray scattering: Establishing links to cuprate superconductors.
\newblock \emph{Phys. Rev. Lett.}, 108:\penalty0 177003, Apr
  2012{\natexlab{b}}.

\bibitem[Kim et~al.(2009)Kim, Ohsumi, Komesu, Sakai, Morita, Takagi, and
  Arima]{Kim2009}
B.~J. Kim, H.~Ohsumi, T.~Komesu, S.~Sakai, T.~Morita, H.~Takagi, and T.~Arima.
\newblock Phase-sensitive observation of a spin-orbital mott state in
  {Sr$_2$IrO$_4$}.
\newblock \emph{Science}, 323\penalty0 (5919):\penalty0 1329--1332, 2009.

\bibitem[Kimel et~al.(2002)Kimel, Pisarev, Hohlfeld, and Rasing]{Kimel2002}
A.~V. Kimel, R.~V. Pisarev, J.~Hohlfeld, and Th. Rasing.
\newblock Ultrafast quenching of the antiferromagnetic order in
  $\mathrm{F}\mathrm{e}\mathrm{B}{\mathrm{o}}_{\mathrm{3}}$: Direct optical
  probing of the phonon-magnon coupling.
\newblock \emph{Phys. Rev. Lett.}, 89:\penalty0 287401, Dec 2002.
\newblock \doi{10.1103/PhysRevLett.89.287401}.
\newblock URL \url{https://link.aps.org/doi/10.1103/PhysRevLett.89.287401}.

\bibitem[Kimel et~al.(2006)Kimel, Stanciu, Usachev, Pisarev, Gridnev, Kirilyuk,
  and Rasing]{Kimel2006}
A.~V. Kimel, C.~D. Stanciu, P.~A. Usachev, R.~V. Pisarev, V.~N. Gridnev,
  A.~Kirilyuk, and Th. Rasing.
\newblock Optical excitation of antiferromagnetic resonance in
  {${\mathrm{TmFeO}}_{3}$}.
\newblock \emph{Phys. Rev. B}, 74:\penalty0 060403, Aug 2006.
\newblock \doi{10.1103/PhysRevB.74.060403}.
\newblock URL \url{https://link.aps.org/doi/10.1103/PhysRevB.74.060403}.

\bibitem[Matsubara et~al.(2007)Matsubara, Okimoto, Ogasawara, Tomioka, Okamoto,
  and Tokura]{Matsubara2007}
M.~Matsubara, Y.~Okimoto, T.~Ogasawara, Y.~Tomioka, H.~Okamoto, and Y.~Tokura.
\newblock Ultrafast photoinduced insulator-ferromagnet transition in the
  perovskite manganite
  {${\mathrm{Gd}}_{0.55}{\mathrm{Sr}}_{0.45}{\mathrm{MnO}}_{3}$}.
\newblock \emph{Phys. Rev. Lett.}, 99:\penalty0 207401, Nov 2007.
\newblock \doi{10.1103/PhysRevLett.99.207401}.
\newblock URL \url{https://link.aps.org/doi/10.1103/PhysRevLett.99.207401}.

\bibitem[Yada et~al.(2011)Yada, Matsubara, Matsuzaki, Yamada, Sawa, and
  Okamoto]{Yada2011}
Hiroyuki Yada, Masakazu Matsubara, Hiroyuki Matsuzaki, Hiroyuki Yamada, Akihito
  Sawa, and Hiroshi Okamoto.
\newblock Discrimination between photodoping- and heat-induced magnetization
  changes in {Nd$_{0.52}$Sr$_{0.48}$MnO$_{3}$} using a heterostructure with
  {SrTiO${}_{3}$}.
\newblock \emph{Phys. Rev. B}, 84:\penalty0 045114, Jul 2011.
\newblock \doi{10.1103/PhysRevB.84.045114}.
\newblock URL \url{https://link.aps.org/doi/10.1103/PhysRevB.84.045114}.

\bibitem[Calder et~al.(2018)Calder, Pajerowski, Stone, and May]{Calder2018}
S.~Calder, D.~M. Pajerowski, M.~B. Stone, and A.~F. May.
\newblock Spin-gap and two-dimensional magnetic excitations in {Sr$_2$IrO$_4$}.
\newblock \emph{Phys. Rev. B}, 98:\penalty0 220402, Dec 2018.

\bibitem[Chu et~al.(2017)Chu, Zhao, de~la Torre, Hogan, Wilson, and
  Hsieh]{Chu2017}
H.~Chu, L.~Zhao, A.~de~la Torre, T.~Hogan, S.~D. Wilson, and D.~Hsieh.
\newblock A charge density wave-like instability in a doped
  spin--orbit-assisted weak mott insulator.
\newblock \emph{Nature Materials}, 16:\penalty0 200 EP --, 01 2017.

\bibitem[Li et~al.(2013)Li, Kong, Qi, Jin, Yuan, DeLong, Schlottmann, and
  Cao]{Li2013}
L.~Li, P.~P. Kong, T.~F. Qi, C.~Q. Jin, S.~J. Yuan, L.~E. DeLong,
  P.~Schlottmann, and G.~Cao.
\newblock Tuning the ${J}_{\mathrm{eff}}=\frac{1}{2}$ insulating state via
  electron doping and pressure in the double-layered iridate
  {Sr${}_{3}$Ir${}_{2}$O${}_{7}$}.
\newblock \emph{Phys. Rev. B}, 87:\penalty0 235127, Jun 2013.

\bibitem[Huang and Lindau(2012)]{Huang2012}
Zhirong Huang and Ingolf Lindau.
\newblock Sacla hard-x-ray compact fel.
\newblock \emph{Nature Photonics}, 6\penalty0 (8):\penalty0 505--506, 2012.
\newblock \doi{10.1038/nphoton.2012.184}.

\bibitem[Tono et~al.(2013)Tono, Togashi, Inubushi, Sato, Katayama, Ogawa,
  Ohashi, Kimura, Takahashi, Takeshita, Tomizawa, Goto, Ishikawa, and
  Yabashi]{Tono2013}
K~Tono, T~Togashi, Y~Inubushi, T~Sato, T~Katayama, K~Ogawa, H~Ohashi, H~Kimura,
  S~Takahashi, K~Takeshita, H~Tomizawa, S~Goto, T~Ishikawa, and M~Yabashi.
\newblock Beamline, experimental stations and photon beam diagnostics for the
  hard x-ray free electron laser of {SACLA}.
\newblock \emph{New Journal of Physics}, 15\penalty0 (8):\penalty0 083035, aug
  2013.
\newblock \doi{10.1088/1367-2630/15/8/083035}.

\bibitem[Kameshima et~al.(2014)Kameshima, Ono, Kudo, Ozaki, Kirihara,
  Kobayashi, Inubushi, Yabashi, Horigome, Holland, Holland, Burt, Murao, and
  Hatsui]{Kameshima2014}
Takashi Kameshima, Shun Ono, Togo Kudo, Kyosuke Ozaki, Yoichi Kirihara, Kazuo
  Kobayashi, Yuichi Inubushi, Makina Yabashi, Toshio Horigome, Andrew Holland,
  Karen Holland, David Burt, Hajime Murao, and Takaki Hatsui.
\newblock Development of an x-ray pixel detector with multi-port charge-coupled
  device for x-ray free-electron laser experiments.
\newblock \emph{Review of Scientific Instruments}, 85\penalty0 (3):\penalty0
  033110, 2014.
\newblock \doi{10.1063/1.4867668}.

\bibitem[Katayama et~al.(2016)Katayama, Owada, Togashi, Ogawa, Karvinen,
  Vartiainen, Eronen, David, Sato, Nakajima, Joti, Yumoto, Ohashi, and
  Yabashi]{Katayama2016}
Tetsuo Katayama, Shigeki Owada, Tadashi Togashi, Kanade Ogawa, Petri Karvinen,
  Ismo Vartiainen, Anni Eronen, Christian David, Takahiro Sato, Kyo Nakajima,
  Yasumasa Joti, Hirokatsu Yumoto, Haruhiko Ohashi, and Makina Yabashi.
\newblock A beam branching method for timing and spectral characterization of
  hard x-ray free-electron lasers.
\newblock \emph{Structural Dynamics}, 3\penalty0 (3):\penalty0 034301, 2016.
\newblock \doi{10.1063/1.4939655}.

\bibitem[Chollet et~al.(2015)Chollet, Alonso-Mori, Cammarata, Damiani, Defever,
  Delor, Feng, Glownia, Langton, Nelson, Ramsey, Robert, Sikorski, Song,
  Stefanescu, Srinivasan, Zhu, Lemke, and Fritz]{Chollet2015}
Matthieu Chollet, Roberto Alonso-Mori, Marco Cammarata, Daniel Damiani, Jim
  Defever, James~T Delor, Yiping Feng, James~M Glownia, J~Brian Langton, Silke
  Nelson, Kelley Ramsey, Aymeric Robert, Marcin Sikorski, Sanghoon Song, Daniel
  Stefanescu, Venkat Srinivasan, Diling Zhu, Henrik~T Lemke, and David~M Fritz.
\newblock The x-ray pump-probe instrument at the linac coherent light source.
\newblock \emph{Journal of synchrotron radiation}, 22\penalty0 (3):\penalty0
  503--507, 05 2015.

\bibitem[Carini et~al.(2016)Carini, Alonso-Mori, Blaj, Caragiulo, Chollet,
  Damiani, Dragone, Feng, Haller, Hart, Hasi, Herbst, Herrmann, Kenney, Lemke,
  Manger, Markovic, Mehta, Nelson, Nishimura, Osier, Pines, Reese, Robert,
  Segal, Sikorski, Song, Thayer, Tomada, Weaver, and Zhu]{Carini2016}
G.~A. Carini, R.~Alonso-Mori, G.~Blaj, P.~Caragiulo, M.~Chollet, D.~Damiani,
  A.~Dragone, Y.~Feng, G.~Haller, P.~Hart, J.~Hasi, R.~Herbst, S.~Herrmann,
  C.~Kenney, H.~Lemke, L.~Manger, B.~Markovic, A.~Mehta, S.~Nelson,
  K.~Nishimura, S.~Osier, J.~Pines, B.~Reese, A.~Robert, J.~Segal, M.~Sikorski,
  S.~Song, J.~Thayer, A.~Tomada, M.~Weaver, and D.~Zhu.
\newblock epix100 camera: Use and applications at lcls.
\newblock \emph{AIP Conference Proceedings}, 1741\penalty0 (1):\penalty0
  040008, 2016.
\newblock \doi{10.1063/1.4952880}.

\end{thebibliography}


\begin{thebibliography}{19}%
\makeatletter
\providecommand \@ifxundefined [1]{%
 \@ifx{#1\undefined}
}%
\providecommand \@ifnum [1]{%
 \ifnum #1\expandafter \@firstoftwo
 \else \expandafter \@secondoftwo
 \fi
}%
\providecommand \@ifx [1]{%
 \ifx #1\expandafter \@firstoftwo
 \else \expandafter \@secondoftwo
 \fi
}%
\providecommand \natexlab [1]{#1}%
\providecommand \enquote  [1]{``#1''}%
\providecommand \bibnamefont  [1]{#1}%
\providecommand \bibfnamefont [1]{#1}%
\providecommand \citenamefont [1]{#1}%
\providecommand \href@noop [0]{\@secondoftwo}%
\providecommand \href [0]{\begingroup \@sanitize@url \@href}%
\providecommand \@href[1]{\@@startlink{#1}\@@href}%
\providecommand \@@href[1]{\endgroup#1\@@endlink}%
\providecommand \@sanitize@url [0]{\catcode `\\12\catcode `\$12\catcode
  `\&12\catcode `\#12\catcode `\^12\catcode `\_12\catcode `\%12\relax}%
\providecommand \@@startlink[1]{}%
\providecommand \@@endlink[0]{}%
\providecommand \url  [0]{\begingroup\@sanitize@url \@url }%
\providecommand \@url [1]{\endgroup\@href {#1}{\urlprefix }}%
\providecommand \urlprefix  [0]{URL }%
\providecommand \Eprint [0]{\href }%
\providecommand \doibase [0]{https://doi.org/}%
\providecommand \selectlanguage [0]{\@gobble}%
\providecommand \bibinfo  [0]{\@secondoftwo}%
\providecommand \bibfield  [0]{\@secondoftwo}%
\providecommand \translation [1]{[#1]}%
\providecommand \BibitemOpen [0]{}%
\providecommand \bibitemStop [0]{}%
\providecommand \bibitemNoStop [0]{.\EOS\space}%
\providecommand \EOS [0]{\spacefactor3000\relax}%
\providecommand \BibitemShut  [1]{\csname bibitem#1\endcsname}%
\let\auto@bib@innerbib\@empty
\bibitem [{\citenamefont {Zeiger}\ \emph {et~al.}(1992)\citenamefont {Zeiger},
  \citenamefont {Vidal}, \citenamefont {Cheng}, \citenamefont {Ippen},
  \citenamefont {Dresselhaus},\ and\ \citenamefont {Dresselhaus}}]{Zeiger1992}%
  \BibitemOpen
  \bibfield  {author} {\bibinfo {author} {\bibfnamefont {H.~J.}\ \bibnamefont
  {Zeiger}}, \bibinfo {author} {\bibfnamefont {J.}~\bibnamefont {Vidal}},
  \bibinfo {author} {\bibfnamefont {T.~K.}\ \bibnamefont {Cheng}}, \bibinfo
  {author} {\bibfnamefont {E.~P.}\ \bibnamefont {Ippen}}, \bibinfo {author}
  {\bibfnamefont {G.}~\bibnamefont {Dresselhaus}},\ and\ \bibinfo {author}
  {\bibfnamefont {M.~S.}\ \bibnamefont {Dresselhaus}},\ }\bibfield  {title}
  {\bibinfo {title} {Theory for displacive excitation of coherent phonons},\
  }\href@noop {} {\bibfield  {journal} {\bibinfo  {journal} {Phys. Rev. B}\
  }\textbf {\bibinfo {volume} {45}},\ \bibinfo {pages} {768} (\bibinfo {year}
  {1992})}\BibitemShut {NoStop}%
\bibitem [{\citenamefont {Feist}\ \emph {et~al.}(2018)\citenamefont {Feist},
  \citenamefont {Rubiano~da Silva}, \citenamefont {Liang}, \citenamefont
  {Ropers},\ and\ \citenamefont {Sch{\"a}fer}}]{feist2018nanoscale}%
  \BibitemOpen
  \bibfield  {author} {\bibinfo {author} {\bibfnamefont {A.}~\bibnamefont
  {Feist}}, \bibinfo {author} {\bibfnamefont {N.}~\bibnamefont {Rubiano~da
  Silva}}, \bibinfo {author} {\bibfnamefont {W.}~\bibnamefont {Liang}},
  \bibinfo {author} {\bibfnamefont {C.}~\bibnamefont {Ropers}},\ and\ \bibinfo
  {author} {\bibfnamefont {S.}~\bibnamefont {Sch{\"a}fer}},\ }\bibfield
  {title} {\bibinfo {title} {Nanoscale diffractive probing of strain dynamics
  in ultrafast transmission electron microscopy},\ }\href@noop {} {\bibfield
  {journal} {\bibinfo  {journal} {Structural Dynamics}\ }\textbf {\bibinfo
  {volume} {5}},\ \bibinfo {pages} {014302} (\bibinfo {year}
  {2018})}\BibitemShut {NoStop}%
\bibitem [{\citenamefont {Lejman}\ \emph {et~al.}(2014)\citenamefont {Lejman},
  \citenamefont {Vaudel}, \citenamefont {Infante}, \citenamefont {Gemeiner},
  \citenamefont {Gusev}, \citenamefont {Dkhil},\ and\ \citenamefont
  {Ruello}}]{lejman2014giant}%
  \BibitemOpen
  \bibfield  {author} {\bibinfo {author} {\bibfnamefont {M.}~\bibnamefont
  {Lejman}}, \bibinfo {author} {\bibfnamefont {G.}~\bibnamefont {Vaudel}},
  \bibinfo {author} {\bibfnamefont {I.~C.}\ \bibnamefont {Infante}}, \bibinfo
  {author} {\bibfnamefont {P.}~\bibnamefont {Gemeiner}}, \bibinfo {author}
  {\bibfnamefont {V.~E.}\ \bibnamefont {Gusev}}, \bibinfo {author}
  {\bibfnamefont {B.}~\bibnamefont {Dkhil}},\ and\ \bibinfo {author}
  {\bibfnamefont {P.}~\bibnamefont {Ruello}},\ }\bibfield  {title} {\bibinfo
  {title} {Giant ultrafast photo-induced shear strain in ferroelectric
  {BiFeO$_3$}},\ }\href@noop {} {\bibfield  {journal} {\bibinfo  {journal}
  {Nature communications}\ }\textbf {\bibinfo {volume} {5}},\ \bibinfo {pages}
  {4301} (\bibinfo {year} {2014})}\BibitemShut {NoStop}%
\bibitem [{\citenamefont {Dean}\ \emph {et~al.}(2016)\citenamefont {Dean},
  \citenamefont {Cao}, \citenamefont {Liu}, \citenamefont {Wall}, \citenamefont
  {Zhu}, \citenamefont {Mankowsky}, \citenamefont {Thampy}, \citenamefont
  {Chen}, \citenamefont {Vale}, \citenamefont {Casa}, \citenamefont {Kim},
  \citenamefont {Said}, \citenamefont {Juhas}, \citenamefont {Alonso-Mori},
  \citenamefont {Glownia}, \citenamefont {Robert}, \citenamefont {Robinson},
  \citenamefont {Sikorski}, \citenamefont {Song}, \citenamefont {Kozina},
  \citenamefont {Lemke}, \citenamefont {Patthey}, \citenamefont {Owada},
  \citenamefont {Katayama}, \citenamefont {Yabashi}, \citenamefont {Tanaka},
  \citenamefont {Togashi}, \citenamefont {Liu}, \citenamefont {Rayan~Serrao},
  \citenamefont {Kim}, \citenamefont {Huber}, \citenamefont {Chang},
  \citenamefont {McMorrow}, \citenamefont {F{\"o}rst},\ and\ \citenamefont
  {Hill}}]{Dean2016}%
  \BibitemOpen
  \bibfield  {author} {\bibinfo {author} {\bibfnamefont {M.~P.~M.}\
  \bibnamefont {Dean}}, \bibinfo {author} {\bibfnamefont {Y.}~\bibnamefont
  {Cao}}, \bibinfo {author} {\bibfnamefont {X.}~\bibnamefont {Liu}}, \bibinfo
  {author} {\bibfnamefont {S.}~\bibnamefont {Wall}}, \bibinfo {author}
  {\bibfnamefont {D.}~\bibnamefont {Zhu}}, \bibinfo {author} {\bibfnamefont
  {R.}~\bibnamefont {Mankowsky}}, \bibinfo {author} {\bibfnamefont
  {V.}~\bibnamefont {Thampy}}, \bibinfo {author} {\bibfnamefont {X.~M.}\
  \bibnamefont {Chen}}, \bibinfo {author} {\bibfnamefont {J.~G.}\ \bibnamefont
  {Vale}}, \bibinfo {author} {\bibfnamefont {D.}~\bibnamefont {Casa}}, \bibinfo
  {author} {\bibfnamefont {J.}~\bibnamefont {Kim}}, \bibinfo {author}
  {\bibfnamefont {A.~H.}\ \bibnamefont {Said}}, \bibinfo {author}
  {\bibfnamefont {P.}~\bibnamefont {Juhas}}, \bibinfo {author} {\bibfnamefont
  {R.}~\bibnamefont {Alonso-Mori}}, \bibinfo {author} {\bibfnamefont {J.~M.}\
  \bibnamefont {Glownia}}, \bibinfo {author} {\bibfnamefont {A.}~\bibnamefont
  {Robert}}, \bibinfo {author} {\bibfnamefont {J.}~\bibnamefont {Robinson}},
  \bibinfo {author} {\bibfnamefont {M.}~\bibnamefont {Sikorski}}, \bibinfo
  {author} {\bibfnamefont {S.}~\bibnamefont {Song}}, \bibinfo {author}
  {\bibfnamefont {M.}~\bibnamefont {Kozina}}, \bibinfo {author} {\bibfnamefont
  {H.}~\bibnamefont {Lemke}}, \bibinfo {author} {\bibfnamefont
  {L.}~\bibnamefont {Patthey}}, \bibinfo {author} {\bibfnamefont
  {S.}~\bibnamefont {Owada}}, \bibinfo {author} {\bibfnamefont
  {T.}~\bibnamefont {Katayama}}, \bibinfo {author} {\bibfnamefont
  {M.}~\bibnamefont {Yabashi}}, \bibinfo {author} {\bibfnamefont
  {Y.}~\bibnamefont {Tanaka}}, \bibinfo {author} {\bibfnamefont
  {T.}~\bibnamefont {Togashi}}, \bibinfo {author} {\bibfnamefont
  {J.}~\bibnamefont {Liu}}, \bibinfo {author} {\bibfnamefont {C.}~\bibnamefont
  {Rayan~Serrao}}, \bibinfo {author} {\bibfnamefont {B.~J.}\ \bibnamefont
  {Kim}}, \bibinfo {author} {\bibfnamefont {L.}~\bibnamefont {Huber}}, \bibinfo
  {author} {\bibfnamefont {C.~L.}\ \bibnamefont {Chang}}, \bibinfo {author}
  {\bibfnamefont {D.~F.}\ \bibnamefont {McMorrow}}, \bibinfo {author}
  {\bibfnamefont {M.}~\bibnamefont {F{\"o}rst}},\ and\ \bibinfo {author}
  {\bibfnamefont {J.~P.}\ \bibnamefont {Hill}},\ }\bibfield  {title} {\bibinfo
  {title} {Ultrafast energy- and momentum-resolved dynamics of magnetic
  correlations in the photo-doped mott insulator {Sr$_2$IrO$_4$}},\ }\href@noop
  {} {\bibfield  {journal} {\bibinfo  {journal} {Nature Materials}\ }\textbf
  {\bibinfo {volume} {15}},\ \bibinfo {pages} {601} (\bibinfo {year}
  {2016})}\BibitemShut {NoStop}%
\bibitem [{\citenamefont {Orfanidis}(2016)}]{Orfanidis2016}%
  \BibitemOpen
  \bibfield  {author} {\bibinfo {author} {\bibfnamefont {S.~J.}\ \bibnamefont
  {Orfanidis}},\ }\bibinfo {title} {Electromagnetic waves and antennas}\
  (\bibinfo  {publisher} {Rutgers University},\ \bibinfo {year} {2016})\ Chap.\
  \bibinfo {chapter} {Reflection and Transmission}, pp.\ \bibinfo {pages}
  {153--186}\BibitemShut {NoStop}%
\bibitem [{\citenamefont {Kriegner}\ \emph {et~al.}(2013)\citenamefont
  {Kriegner}, \citenamefont {Wintersberger},\ and\ \citenamefont
  {Stangl}}]{Kriegner2013}%
  \BibitemOpen
  \bibfield  {author} {\bibinfo {author} {\bibfnamefont {D.}~\bibnamefont
  {Kriegner}}, \bibinfo {author} {\bibfnamefont {E.}~\bibnamefont
  {Wintersberger}},\ and\ \bibinfo {author} {\bibfnamefont {J.}~\bibnamefont
  {Stangl}},\ }\bibfield  {title} {\bibinfo {title} {{{\it xrayutilities}: a
  versatile tool for reciprocal space conversion of scattering data recorded
  with linear and area detectors}},\ }\href
  {https://doi.org/10.1107/S0021889813017214} {\bibfield  {journal} {\bibinfo
  {journal} {Journal of Applied Crystallography}\ }\textbf {\bibinfo {volume}
  {46}},\ \bibinfo {pages} {1162} (\bibinfo {year} {2013})}\BibitemShut
  {NoStop}%
\bibitem [{\citenamefont {Donnerer}\ \emph {et~al.}(2018)\citenamefont
  {Donnerer}, \citenamefont {Sala}, \citenamefont {Pascarelli}, \citenamefont
  {Rosa}, \citenamefont {Andreev}, \citenamefont {Mazurenko}, \citenamefont
  {Irifune}, \citenamefont {Hunter}, \citenamefont {Perry},\ and\ \citenamefont
  {McMorrow}}]{Donnerer2018high}%
  \BibitemOpen
  \bibfield  {author} {\bibinfo {author} {\bibfnamefont {C.}~\bibnamefont
  {Donnerer}}, \bibinfo {author} {\bibfnamefont {M.~M.}\ \bibnamefont {Sala}},
  \bibinfo {author} {\bibfnamefont {S.}~\bibnamefont {Pascarelli}}, \bibinfo
  {author} {\bibfnamefont {A.~D.}\ \bibnamefont {Rosa}}, \bibinfo {author}
  {\bibfnamefont {S.~N.}\ \bibnamefont {Andreev}}, \bibinfo {author}
  {\bibfnamefont {V.~V.}\ \bibnamefont {Mazurenko}}, \bibinfo {author}
  {\bibfnamefont {T.}~\bibnamefont {Irifune}}, \bibinfo {author} {\bibfnamefont
  {E.~C.}\ \bibnamefont {Hunter}}, \bibinfo {author} {\bibfnamefont {R.~S.}\
  \bibnamefont {Perry}},\ and\ \bibinfo {author} {\bibfnamefont {D.~F.}\
  \bibnamefont {McMorrow}},\ }\bibfield  {title} {\bibinfo {title}
  {High-pressure insulator-to-metal transition in
  {${\mathrm{Sr}}_{3}{\mathrm{Ir}}_{2}{\mathrm{O}}_{7}$} studied by x-ray
  absorption spectroscopy},\ }\href
  {https://doi.org/10.1103/PhysRevB.97.035106} {\bibfield  {journal} {\bibinfo
  {journal} {Phys. Rev. B}\ }\textbf {\bibinfo {volume} {97}},\ \bibinfo
  {pages} {035106} (\bibinfo {year} {2018})}\BibitemShut {NoStop}%
\bibitem [{\citenamefont {Ahn}\ \emph {et~al.}(2016)\citenamefont {Ahn},
  \citenamefont {Song}, \citenamefont {Hogan}, \citenamefont {Wilson},\ and\
  \citenamefont {Moon}}]{Ahn2016}%
  \BibitemOpen
  \bibfield  {author} {\bibinfo {author} {\bibfnamefont {G.}~\bibnamefont
  {Ahn}}, \bibinfo {author} {\bibfnamefont {S.~J.}\ \bibnamefont {Song}},
  \bibinfo {author} {\bibfnamefont {T.}~\bibnamefont {Hogan}}, \bibinfo
  {author} {\bibfnamefont {S.~D.}\ \bibnamefont {Wilson}},\ and\ \bibinfo
  {author} {\bibfnamefont {S.~J.}\ \bibnamefont {Moon}},\ }\bibfield  {title}
  {\bibinfo {title} {Infrared spectroscopic evidences of strong electronic
  correlations in {(Sr$_{1-x}$La$_x$)$_3$Ir$_2$O$_7$}},\ }\href
  {https://doi.org/10.1038/srep32632} {\bibfield  {journal} {\bibinfo
  {journal} {Scientific Reports}\ }\textbf {\bibinfo {volume} {6}},\ \bibinfo
  {pages} {32632} (\bibinfo {year} {2016})}\BibitemShut {NoStop}%
\bibitem [{\citenamefont {Jackeli}\ and\ \citenamefont
  {Khaliullin}(2009)}]{Jackeli2009}%
  \BibitemOpen
  \bibfield  {author} {\bibinfo {author} {\bibfnamefont {G.}~\bibnamefont
  {Jackeli}}\ and\ \bibinfo {author} {\bibfnamefont {G.}~\bibnamefont
  {Khaliullin}},\ }\bibfield  {title} {\bibinfo {title} {Mott insulators in the
  strong spin-orbit coupling limit: From {Heisenberg} to a quantum compass and
  {Kitaev} models},\ }\href {https://doi.org/10.1103/PhysRevLett.102.017205}
  {\bibfield  {journal} {\bibinfo  {journal} {Phys. Rev. Lett.}\ }\textbf
  {\bibinfo {volume} {102}},\ \bibinfo {pages} {017205} (\bibinfo {year}
  {2009})}\BibitemShut {NoStop}%
\bibitem [{\citenamefont {Rau}\ \emph {et~al.}(2016)\citenamefont {Rau},
  \citenamefont {Lee},\ and\ \citenamefont {Kee}}]{Jeffrey2016}%
  \BibitemOpen
  \bibfield  {author} {\bibinfo {author} {\bibfnamefont {J.~G.}\ \bibnamefont
  {Rau}}, \bibinfo {author} {\bibfnamefont {E.~K.-H.}\ \bibnamefont {Lee}},\
  and\ \bibinfo {author} {\bibfnamefont {H.-Y.}\ \bibnamefont {Kee}},\
  }\bibfield  {title} {\bibinfo {title} {Spin-orbit physics giving rise to
  novel phases in correlated systems: Iridates and related materials},\
  }\href@noop {} {\bibfield  {journal} {\bibinfo  {journal} {Annual Review of
  Condensed Matter Physics}\ }\textbf {\bibinfo {volume} {7}},\ \bibinfo
  {pages} {195} (\bibinfo {year} {2016})}\BibitemShut {NoStop}%
\bibitem [{\citenamefont {Rayan~Serrao}\ \emph {et~al.}(2013)\citenamefont
  {Rayan~Serrao}, \citenamefont {Liu}, \citenamefont {Heron}, \citenamefont
  {Singh-Bhalla}, \citenamefont {Yadav}, \citenamefont {Suresha}, \citenamefont
  {Paull}, \citenamefont {Yi}, \citenamefont {Chu}, \citenamefont {Trassin},
  \citenamefont {Vishwanath}, \citenamefont {Arenholz}, \citenamefont
  {Frontera}, \citenamefont {\ifmmode~\check{Z}\else \v{Z}\fi{}elezn\'y},
  \citenamefont {Jungwirth}, \citenamefont {Marti},\ and\ \citenamefont
  {Ramesh}}]{Serrao2013}%
  \BibitemOpen
  \bibfield  {author} {\bibinfo {author} {\bibfnamefont {C.}~\bibnamefont
  {Rayan~Serrao}}, \bibinfo {author} {\bibfnamefont {J.}~\bibnamefont {Liu}},
  \bibinfo {author} {\bibfnamefont {J.~T.}\ \bibnamefont {Heron}}, \bibinfo
  {author} {\bibfnamefont {G.}~\bibnamefont {Singh-Bhalla}}, \bibinfo {author}
  {\bibfnamefont {A.}~\bibnamefont {Yadav}}, \bibinfo {author} {\bibfnamefont
  {S.~J.}\ \bibnamefont {Suresha}}, \bibinfo {author} {\bibfnamefont {R.~J.}\
  \bibnamefont {Paull}}, \bibinfo {author} {\bibfnamefont {D.}~\bibnamefont
  {Yi}}, \bibinfo {author} {\bibfnamefont {J.-H.}\ \bibnamefont {Chu}},
  \bibinfo {author} {\bibfnamefont {M.}~\bibnamefont {Trassin}}, \bibinfo
  {author} {\bibfnamefont {A.}~\bibnamefont {Vishwanath}}, \bibinfo {author}
  {\bibfnamefont {E.}~\bibnamefont {Arenholz}}, \bibinfo {author}
  {\bibfnamefont {C.}~\bibnamefont {Frontera}}, \bibinfo {author}
  {\bibfnamefont {J.}~\bibnamefont {\ifmmode~\check{Z}\else
  \v{Z}\fi{}elezn\'y}}, \bibinfo {author} {\bibfnamefont {T.}~\bibnamefont
  {Jungwirth}}, \bibinfo {author} {\bibfnamefont {X.}~\bibnamefont {Marti}},\
  and\ \bibinfo {author} {\bibfnamefont {R.}~\bibnamefont {Ramesh}},\
  }\bibfield  {title} {\bibinfo {title} {Epitaxy-distorted spin-orbit {Mott}
  insulator in {Sr$_{2}$IrO$_{4}$} thin films},\ }\href
  {https://doi.org/10.1103/PhysRevB.87.085121} {\bibfield  {journal} {\bibinfo
  {journal} {Phys. Rev. B}\ }\textbf {\bibinfo {volume} {87}},\ \bibinfo
  {pages} {085121} (\bibinfo {year} {2013})}\BibitemShut {NoStop}%
\bibitem [{\citenamefont {Hogan}\ \emph {et~al.}(2016)\citenamefont {Hogan},
  \citenamefont {Bjaalie}, \citenamefont {Zhao}, \citenamefont {Belvin},
  \citenamefont {Wang}, \citenamefont {Van~de Walle}, \citenamefont {Hsieh},\
  and\ \citenamefont {Wilson}}]{Hogan2016}%
  \BibitemOpen
  \bibfield  {author} {\bibinfo {author} {\bibfnamefont {T.}~\bibnamefont
  {Hogan}}, \bibinfo {author} {\bibfnamefont {L.}~\bibnamefont {Bjaalie}},
  \bibinfo {author} {\bibfnamefont {L.}~\bibnamefont {Zhao}}, \bibinfo {author}
  {\bibfnamefont {C.}~\bibnamefont {Belvin}}, \bibinfo {author} {\bibfnamefont
  {X.}~\bibnamefont {Wang}}, \bibinfo {author} {\bibfnamefont {C.~G.}\
  \bibnamefont {Van~de Walle}}, \bibinfo {author} {\bibfnamefont
  {D.}~\bibnamefont {Hsieh}},\ and\ \bibinfo {author} {\bibfnamefont {S.~D.}\
  \bibnamefont {Wilson}},\ }\bibfield  {title} {\bibinfo {title} {Structural
  investigation of the bilayer iridate
  {${\mathrm{Sr}}_{3}{\mathrm{Ir}}_{2}{\mathrm{O}}_{7}$}},\ }\href
  {https://doi.org/10.1103/PhysRevB.93.134110} {\bibfield  {journal} {\bibinfo
  {journal} {Phys. Rev. B}\ }\textbf {\bibinfo {volume} {93}},\ \bibinfo
  {pages} {134110} (\bibinfo {year} {2016})}\BibitemShut {NoStop}%
\bibitem [{\citenamefont {Takayama}\ \emph {et~al.}(2016)\citenamefont
  {Takayama}, \citenamefont {Matsumoto}, \citenamefont {Jackeli},\ and\
  \citenamefont {Takagi}}]{Takayama2016}%
  \BibitemOpen
  \bibfield  {author} {\bibinfo {author} {\bibfnamefont {T.}~\bibnamefont
  {Takayama}}, \bibinfo {author} {\bibfnamefont {A.}~\bibnamefont {Matsumoto}},
  \bibinfo {author} {\bibfnamefont {G.}~\bibnamefont {Jackeli}},\ and\ \bibinfo
  {author} {\bibfnamefont {H.}~\bibnamefont {Takagi}},\ }\bibfield  {title}
  {\bibinfo {title} {Model analysis of magnetic susceptibility of
  {Sr$_{2}$IrO$_4$}: A two-dimensional ${J}_{\mathrm{eff}}=\frac{1}{2}$
  {Heisenberg} system with competing interlayer couplings},\ }\href
  {https://doi.org/10.1103/PhysRevB.94.224420} {\bibfield  {journal} {\bibinfo
  {journal} {Phys. Rev. B}\ }\textbf {\bibinfo {volume} {94}},\ \bibinfo
  {pages} {224420} (\bibinfo {year} {2016})}\BibitemShut {NoStop}%
\bibitem [{\citenamefont {Liu}\ and\ \citenamefont
  {Khaliullin}(2019)}]{Liu2019}%
  \BibitemOpen
  \bibfield  {author} {\bibinfo {author} {\bibfnamefont {H.}~\bibnamefont
  {Liu}}\ and\ \bibinfo {author} {\bibfnamefont {G.}~\bibnamefont
  {Khaliullin}},\ }\bibfield  {title} {\bibinfo {title} {Pseudo-{Jahn-Teller}
  effect and magnetoelastic coupling in spin-orbit {Mott} insulators},\ }\href
  {https://doi.org/10.1103/PhysRevLett.122.057203} {\bibfield  {journal}
  {\bibinfo  {journal} {Phys. Rev. Lett.}\ }\textbf {\bibinfo {volume} {122}},\
  \bibinfo {pages} {057203} (\bibinfo {year} {2019})}\BibitemShut {NoStop}%
\bibitem [{\citenamefont {Porras}\ \emph {et~al.}(2019)\citenamefont {Porras},
  \citenamefont {Bertinshaw}, \citenamefont {Liu}, \citenamefont {Khaliullin},
  \citenamefont {Sung}, \citenamefont {Kim}, \citenamefont {Francoual},
  \citenamefont {Steffens}, \citenamefont {Deng}, \citenamefont {Sala},
  \citenamefont {Efimenko}, \citenamefont {Said}, \citenamefont {Casa},
  \citenamefont {Huang}, \citenamefont {Gog}, \citenamefont {Kim},
  \citenamefont {Keimer},\ and\ \citenamefont {Kim}}]{Porras2019}%
  \BibitemOpen
  \bibfield  {author} {\bibinfo {author} {\bibfnamefont {J.}~\bibnamefont
  {Porras}}, \bibinfo {author} {\bibfnamefont {J.}~\bibnamefont {Bertinshaw}},
  \bibinfo {author} {\bibfnamefont {H.}~\bibnamefont {Liu}}, \bibinfo {author}
  {\bibfnamefont {G.}~\bibnamefont {Khaliullin}}, \bibinfo {author}
  {\bibfnamefont {N.~H.}\ \bibnamefont {Sung}}, \bibinfo {author}
  {\bibfnamefont {J.-W.}\ \bibnamefont {Kim}}, \bibinfo {author} {\bibfnamefont
  {S.}~\bibnamefont {Francoual}}, \bibinfo {author} {\bibfnamefont
  {P.}~\bibnamefont {Steffens}}, \bibinfo {author} {\bibfnamefont
  {G.}~\bibnamefont {Deng}}, \bibinfo {author} {\bibfnamefont {M.~M.}\
  \bibnamefont {Sala}}, \bibinfo {author} {\bibfnamefont {A.}~\bibnamefont
  {Efimenko}}, \bibinfo {author} {\bibfnamefont {A.}~\bibnamefont {Said}},
  \bibinfo {author} {\bibfnamefont {D.}~\bibnamefont {Casa}}, \bibinfo {author}
  {\bibfnamefont {X.}~\bibnamefont {Huang}}, \bibinfo {author} {\bibfnamefont
  {T.}~\bibnamefont {Gog}}, \bibinfo {author} {\bibfnamefont {J.}~\bibnamefont
  {Kim}}, \bibinfo {author} {\bibfnamefont {B.}~\bibnamefont {Keimer}},\ and\
  \bibinfo {author} {\bibfnamefont {B.~J.}\ \bibnamefont {Kim}},\ }\bibfield
  {title} {\bibinfo {title} {Pseudospin-lattice coupling in the spin-orbit
  {Mott} insulator {${\mathrm{Sr}}_{2}{\mathrm{IrO}}_{4}$}},\ }\href
  {https://doi.org/10.1103/PhysRevB.99.085125} {\bibfield  {journal} {\bibinfo
  {journal} {Phys. Rev. B}\ }\textbf {\bibinfo {volume} {99}},\ \bibinfo
  {pages} {085125} (\bibinfo {year} {2019})}\BibitemShut {NoStop}%
\bibitem [{\citenamefont {Kim}\ \emph {et~al.}(2012{\natexlab{a}})\citenamefont
  {Kim}, \citenamefont {Casa}, \citenamefont {Upton}, \citenamefont {Gog},
  \citenamefont {Kim}, \citenamefont {Mitchell}, \citenamefont {van
  Veenendaal}, \citenamefont {Daghofer}, \citenamefont {van~den Brink},
  \citenamefont {Khaliullin},\ and\ \citenamefont {Kim}}]{Kim2012}%
  \BibitemOpen
  \bibfield  {author} {\bibinfo {author} {\bibfnamefont {J.}~\bibnamefont
  {Kim}}, \bibinfo {author} {\bibfnamefont {D.}~\bibnamefont {Casa}}, \bibinfo
  {author} {\bibfnamefont {M.~H.}\ \bibnamefont {Upton}}, \bibinfo {author}
  {\bibfnamefont {T.}~\bibnamefont {Gog}}, \bibinfo {author} {\bibfnamefont
  {Y.-J.}\ \bibnamefont {Kim}}, \bibinfo {author} {\bibfnamefont {J.~F.}\
  \bibnamefont {Mitchell}}, \bibinfo {author} {\bibfnamefont {M.}~\bibnamefont
  {van Veenendaal}}, \bibinfo {author} {\bibfnamefont {M.}~\bibnamefont
  {Daghofer}}, \bibinfo {author} {\bibfnamefont {J.}~\bibnamefont {van~den
  Brink}}, \bibinfo {author} {\bibfnamefont {G.}~\bibnamefont {Khaliullin}},\
  and\ \bibinfo {author} {\bibfnamefont {B.~J.}\ \bibnamefont {Kim}},\
  }\bibfield  {title} {\bibinfo {title} {Magnetic excitation spectra of
  {Sr$_2$IrO$_4$}: Probed by resonant inelastic x-ray scattering: Establishing
  links to cuprate superconductors},\ }\href@noop {} {\bibfield  {journal}
  {\bibinfo  {journal} {Phys. Rev. Lett.}\ }\textbf {\bibinfo {volume} {108}},\
  \bibinfo {pages} {177003} (\bibinfo {year} {2012}{\natexlab{a}})}\BibitemShut
  {NoStop}%
\bibitem [{\citenamefont {Kim}\ \emph {et~al.}(2012{\natexlab{b}})\citenamefont
  {Kim}, \citenamefont {Said}, \citenamefont {Casa}, \citenamefont {Upton},
  \citenamefont {Gog}, \citenamefont {Daghofer}, \citenamefont {Jackeli},
  \citenamefont {van~den Brink}, \citenamefont {Khaliullin},\ and\
  \citenamefont {Kim}}]{Kim2012_2}%
  \BibitemOpen
  \bibfield  {author} {\bibinfo {author} {\bibfnamefont {J.}~\bibnamefont
  {Kim}}, \bibinfo {author} {\bibfnamefont {A.~H.}\ \bibnamefont {Said}},
  \bibinfo {author} {\bibfnamefont {D.}~\bibnamefont {Casa}}, \bibinfo {author}
  {\bibfnamefont {M.~H.}\ \bibnamefont {Upton}}, \bibinfo {author}
  {\bibfnamefont {T.}~\bibnamefont {Gog}}, \bibinfo {author} {\bibfnamefont
  {M.}~\bibnamefont {Daghofer}}, \bibinfo {author} {\bibfnamefont
  {G.}~\bibnamefont {Jackeli}}, \bibinfo {author} {\bibfnamefont
  {J.}~\bibnamefont {van~den Brink}}, \bibinfo {author} {\bibfnamefont
  {G.}~\bibnamefont {Khaliullin}},\ and\ \bibinfo {author} {\bibfnamefont
  {B.~J.}\ \bibnamefont {Kim}},\ }\bibfield  {title} {\bibinfo {title} {Large
  spin-wave energy gap in the bilayer iridate {Sr$_3$Ir$_2$O$_7$}: Evidence for
  enhanced dipolar interactions near the {Mott} metal-insulator transition},\
  }\href@noop {} {\bibfield  {journal} {\bibinfo  {journal} {Phys. Rev. Lett.}\
  }\textbf {\bibinfo {volume} {109}},\ \bibinfo {pages} {157402} (\bibinfo
  {year} {2012}{\natexlab{b}})}\BibitemShut {NoStop}%
\bibitem [{\citenamefont {Gretarsson}\ \emph {et~al.}(2016)\citenamefont
  {Gretarsson}, \citenamefont {Sung}, \citenamefont {H\"oppner}, \citenamefont
  {Kim}, \citenamefont {Keimer},\ and\ \citenamefont
  {Le~Tacon}}]{Gretarsson2015}%
  \BibitemOpen
  \bibfield  {author} {\bibinfo {author} {\bibfnamefont {H.}~\bibnamefont
  {Gretarsson}}, \bibinfo {author} {\bibfnamefont {N.~H.}\ \bibnamefont
  {Sung}}, \bibinfo {author} {\bibfnamefont {M.}~\bibnamefont {H\"oppner}},
  \bibinfo {author} {\bibfnamefont {B.~J.}\ \bibnamefont {Kim}}, \bibinfo
  {author} {\bibfnamefont {B.}~\bibnamefont {Keimer}},\ and\ \bibinfo {author}
  {\bibfnamefont {M.}~\bibnamefont {Le~Tacon}},\ }\bibfield  {title} {\bibinfo
  {title} {Two-magnon {Raman} scattering and pseudospin-lattice interactions in
  {${\mathrm{Sr}}_{2}{\mathrm{IrO}}_{4}$} and
  {${\mathrm{Sr}}_{3}{\mathrm{Ir}}_{2}{\mathrm{O}}_{7}$}},\ }\href
  {https://doi.org/10.1103/PhysRevLett.116.136401} {\bibfield  {journal}
  {\bibinfo  {journal} {Phys. Rev. Lett.}\ }\textbf {\bibinfo {volume} {116}},\
  \bibinfo {pages} {136401} (\bibinfo {year} {2016})}\BibitemShut {NoStop}%
\bibitem [{\citenamefont {Irkhin}\ and\ \citenamefont
  {Katanin}(1997)}]{Irkhin1997}%
  \BibitemOpen
  \bibfield  {author} {\bibinfo {author} {\bibfnamefont {V.~Y.}\ \bibnamefont
  {Irkhin}}\ and\ \bibinfo {author} {\bibfnamefont {A.~A.}\ \bibnamefont
  {Katanin}},\ }\bibfield  {title} {\bibinfo {title} {Critical behavior and the
  n\'eel temperature of quantum quasi-two-dimensional heisenberg
  antiferromagnets},\ }\href@noop {} {\bibfield  {journal} {\bibinfo  {journal}
  {Phys. Rev. B}\ }\textbf {\bibinfo {volume} {55}},\ \bibinfo {pages} {12318}
  (\bibinfo {year} {1997})}\BibitemShut {NoStop}%
\end{thebibliography}%

\end{document}


\title{Supporting Information (SI) Appendix\\Laser-Induced Transient Magnons in Sr$_3$Ir$_2$O$_7$ Throughout the Brillouin Zone}
\renewcommand{\thefigure}{S\arabic{figure}}

\maketitle
\section{SI time-resolved resonant elastic X-ray scattering results}
\begin{figure*}[tbh]
\includegraphics[width=0.7\linewidth]{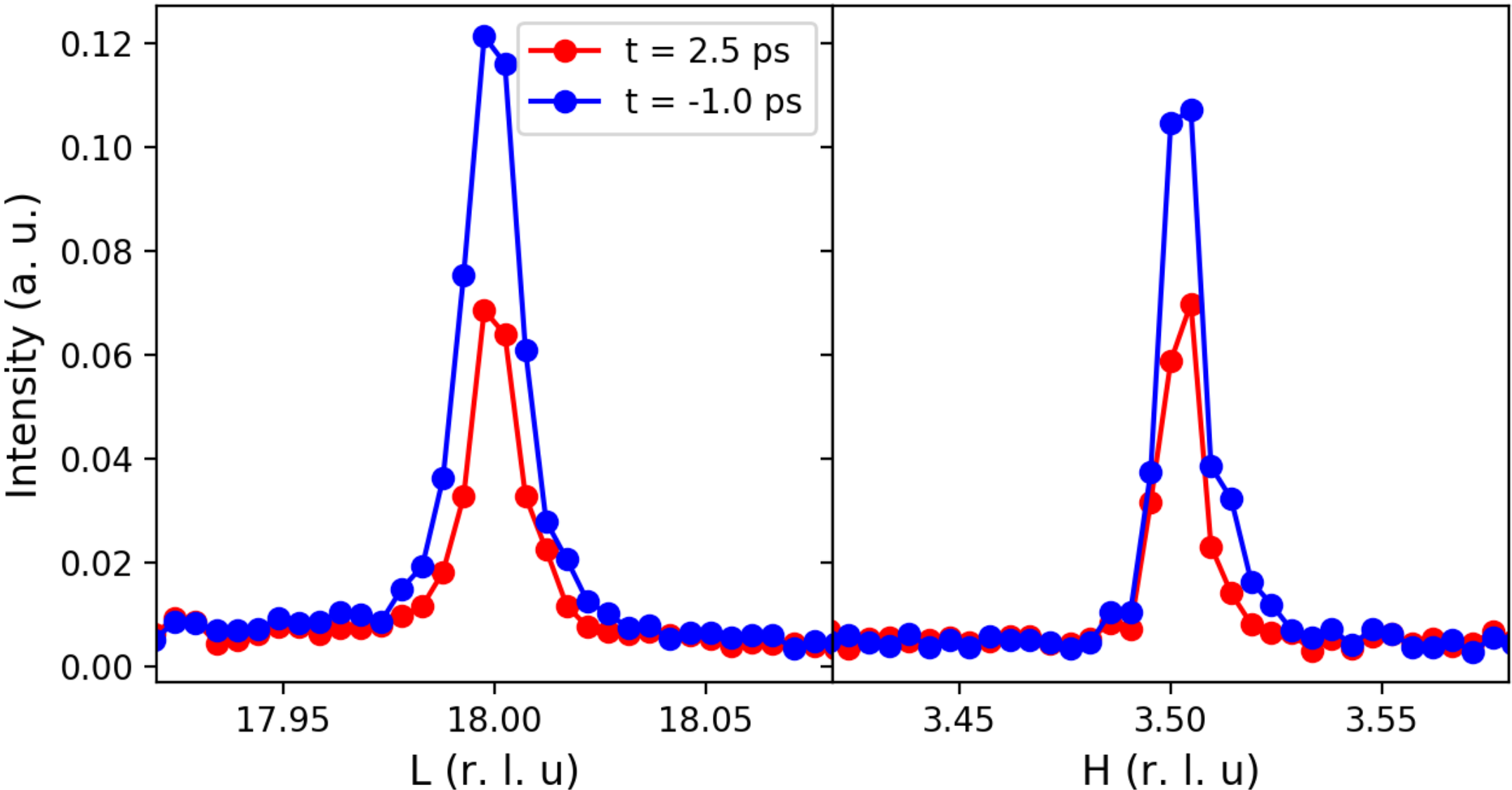}
\caption{Laser effect on the magnetic Bragg peak. Reciprocal lattice scans along (0, 0, $L$) and ($H$, 0, 0) around (-3.5, 1.5, 18) magnetic Bragg peak in reciprocal lattice units (r.l.u). The blue color denotes a scan that was taken before, and the red dataset $\sim$2.5~ps after the arrival of a 32.2~mJ/cm$^2$ strong laser pulse. Only a suppression of the Bragg peak is observed.}
\label{SBragg_peak}
\end{figure*}

The response of magnetic long-range order on the laser pump was first investigated via reciprocal lattice scans along the (0, 0, $L$) and ($H$, 0, 0) directions around the (-3.5, 1.5, 18) magnetic Bragg peak [notation in reciprocal lattice units (r.l.u)]. Figure~\ref{SBragg_peak} displays datasets before and $\sim$2.5~ps after the arrival of a 32.2~mJ/cm$^2$ strong laser pulse. The laser radiation induces a suppression, but no broadening, of the magnetic Bragg peak within our experimental $Q$-resolution. This provides evidence that all crucial information on transient magnetic long-range order is captured by the evolution of the magnetic Bragg peak intensity. It is noted that additional $H$- and $L$-scans were taken $\sim$40~ps after laser excitation confirming this conclusion.

\begin{figure*}[tbh]
\includegraphics[width=\textwidth]{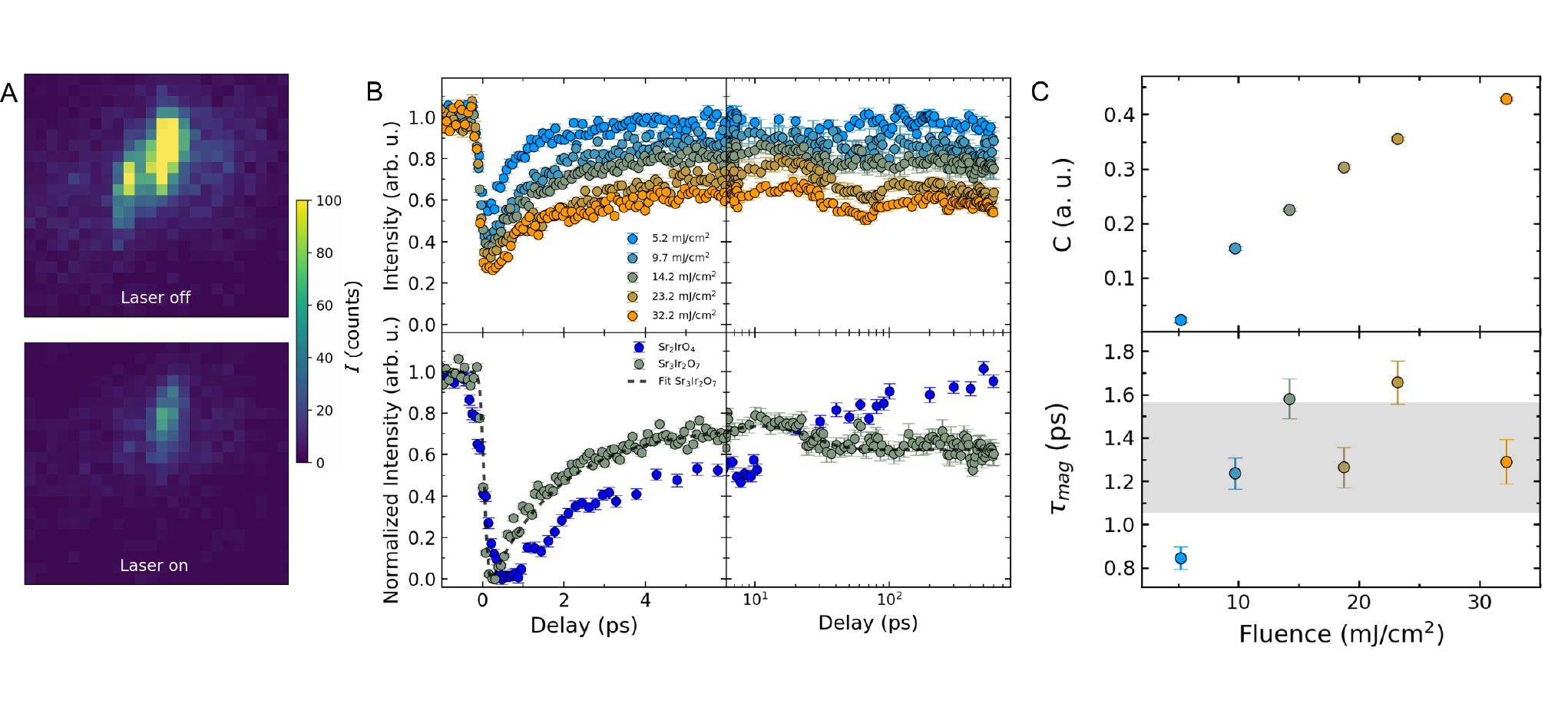} %
\caption{Time evolution of magnetic order in Sr$_3$Ir$_2$O$_7$. (A) Intensity of the (-3.5, 1.5, 18) magnetic Bragg peak [notation in reciprocal lattice units (r.l.u)] in equilibrium (top) and after laser radiation of $\sim$20~mJ/cm$^2$ (bottom). The detector pixels are 0.0086~r.l.u wide (B) Top panel: Relative magnetic Bragg peak intensity as a function of time delay. The error bars follow Poissonian statistics. Panel is split into short (6~ps) and long (100~ps) time ranges for clarity. Bottom panel:  14.2~mJ/cm$^2$ Sr$_3$Ir$_2$O$_7$ data normalized to the maximal depleted volume fraction and fit with the model shown in Eq.~\ref{model}. The results are compared to data for Sr$_2$IrO$_4$  taken from Ref. \cite{Dean2016} with a fluence chosen to match the fast recovery of the magnetic volume fraction. Heisenberg-like Sr$_2$IrO$_4$ reveals a slow recovery with onset around 100~ps, which is absent in gapped Sr$_3$Ir$_2$O$_7$. (C) Fluence dependence of the persistent transient fraction, $C$, and the fast recovery timescale $\tau_\text{mag}$. The gray area shows the standard deviation around the statistical mean value of the fast recovery timescale, $i.e.$ $\tau_\text{mag}$ = 1.3(3)~ps.
}
\label{SREXS_evolution}
\end{figure*}

Figure~\ref{SREXS_evolution}A shows the (-3.5, 1.5, 18) magnetic Bragg signal on the area detector before and around 1~ps after the arrival of the laser pulse. The time evolution of the depleted magnetic Bragg peak intensity is shown in the upper panel of Fig.~\ref{SREXS_evolution}B. Immediately after the optical pump ($t$ = 0) magnetism is reduced by 50-75\% for laser fluences between 5.2 and 32.2~mJ/cm$^2$. A significant fraction of the magnetic signal is recovered within the first picoseconds, but a slower process is superimposed preventing the full restoration of the original magnetic state. As a result, the reduced magnetic peak intensity stays roughly constant in the range between 5 and 500~ps without noticeable recovery. Furthermore, it is noted that a faint oscillatory behavior appears $\sim$50~ps after laser impact for fluences larger than 14.2~mJ/cm$^2$. This feature is a typical signature of a strain wave that is propagating through the crystal and subsequently reflected at the sample boundary \cite{Zeiger1992,feist2018nanoscale,lejman2014giant}.

The time evolution of the transient magnetic long-range order is described by the minimal phenomenological model
\begin{equation}
I(t)=
  \begin{cases}
    1 - A e^{-t/\tau_\text{mag}} - C (1-e^{-t/\tau_{\text{mag}}})  + D e^{-t/\tau_\text{wave}}  \sin(\omega t), & \text{for } t \geq 0 \\
    1                   , & \text{for } t < 0 
    \end{cases},\label{model}
\end{equation}
where $A$ quantifies a prompt, step-like decay at $t = 0$ (as it is much faster than the time resolution), which recovers on a timescale $\tau_\text{mag}$, and $C$ is the fraction of the order that has not been restored in 500~ps. The strain wave features an amplitude $D$, decay time $\tau_\text{wave}$ and oscillation frequency $\omega$. The measured magnetic Bragg peak intensity is modeled by convolving $I(t)$ with a Gaussian function in order to account for our finite time resolution of 0.15~ps (full-width at half-maximum). The initial decay of magnetic long-range order was found to be much faster than the time resolution, and was thus approximated as an instantaneous step function in Eq.~\ref{model}. 

Robust fits of experimental results were obtained by constraining the amplitude of the strain wave to $D$ = 0.5. The frequency of the strain wave at 23.3 mJ/cm$^2$ was fixed to 0.03~rad/s (the mean value of all other conditions) to further stabilize the fit.  All other parameters were varied freely and shown in Tab.~1. It is noted that even for strong fluences $A$ $<$ 1, suggesting some experimental mismatch between pumped laser and probed x-ray volumes (see also below). This emphasizes the need to also include contributions like the energy transfer out of the illuminated sample volume in theoretical approaches that model transient magnetic states.

The lower panel of Fig.~\ref{SREXS_evolution}B displays the best fit to the normalized intensity at 14.2 mJ/cm$^2$ (see also Materials and Methods section). The fluence dependence of $\tau_\text{mag}$ and $C$ is displayed in Fig.~\ref{SREXS_evolution}C. The fast timescale $\tau_\text{mag}$ = 1.3(3)~ps is fluence independent and we find $C$ $>$ 0 for all fluences. This is different than in nearly gapless Sr$_2$IrO$_4$ - a related material where at fluences leading to a similar fast recovery of the magnetic volume fraction, the onset of a slow magnetic recovery is observed around 100 ps (see lower panel of Fig.~\ref{SREXS_evolution}B) \cite{Dean2016}. The different behaviors are likely to arise from their distinct magnetic interactions and resulting magnon dispersions. A phonon-assisted energy transfer from excited (pseudo-)spins into the lattice degrees of freedom, for instance, is expected to be reduced in large gap antiferromagnets, yielding a much slower magnetic recovery when compared to nearly gapless materials.

\begin{table}[b]
\begin{ruledtabular}
\caption{Time evolution of REXS fitting parameters. Time dependence of the depleted magnetic volume fraction $A$, the unrecovered volume $C$ after 500~ps, the fast recovery timescale $\tau_\text{mag}$ and the timescale $\tau_\text{wave}$ and frequency $\omega$ of the strain wave.}
\label{Tab1}
\begin{tabular}{c c c c c c}
  fluence (mJ/cm$^2$) & $A$ & $C$ & $\tau_\text{mag}$ (ps) & $\tau_\text{wave}$ (ps) & $\omega$ (rad/s) \\
  \hline
    5.2 & 0.54(4) & 0.023(5) & 0.85(5) & NaN & NaN \\
  9.7 & 0.66(3) & 0.154(4) & 1.24(7) & 12(2) & 0.025(4)\\
  14.2 & 0.65(2) & 0.225(3) & 1.58(9) & 11(1) & 0.035(4)\\
  18.7 & 0.72(4) & 0.303(4) & 1.26(9) & 18(2) & 0.027(3)\\
  23.2 & 0.71(2) & 0.356(4) & 1.7(1) & 19(1) & 0.03 (fixed)\\
  32.2 & 0.78(2) & 0.429(4) & 1.3(1) & 13(1) & 0.033(4)\\
\end{tabular}
\end{ruledtabular}
\end{table}
\section{SI time-resolved resonant inelastic X-ray scattering results}

\begin{figure*}[t!]
\includegraphics[width=\linewidth]{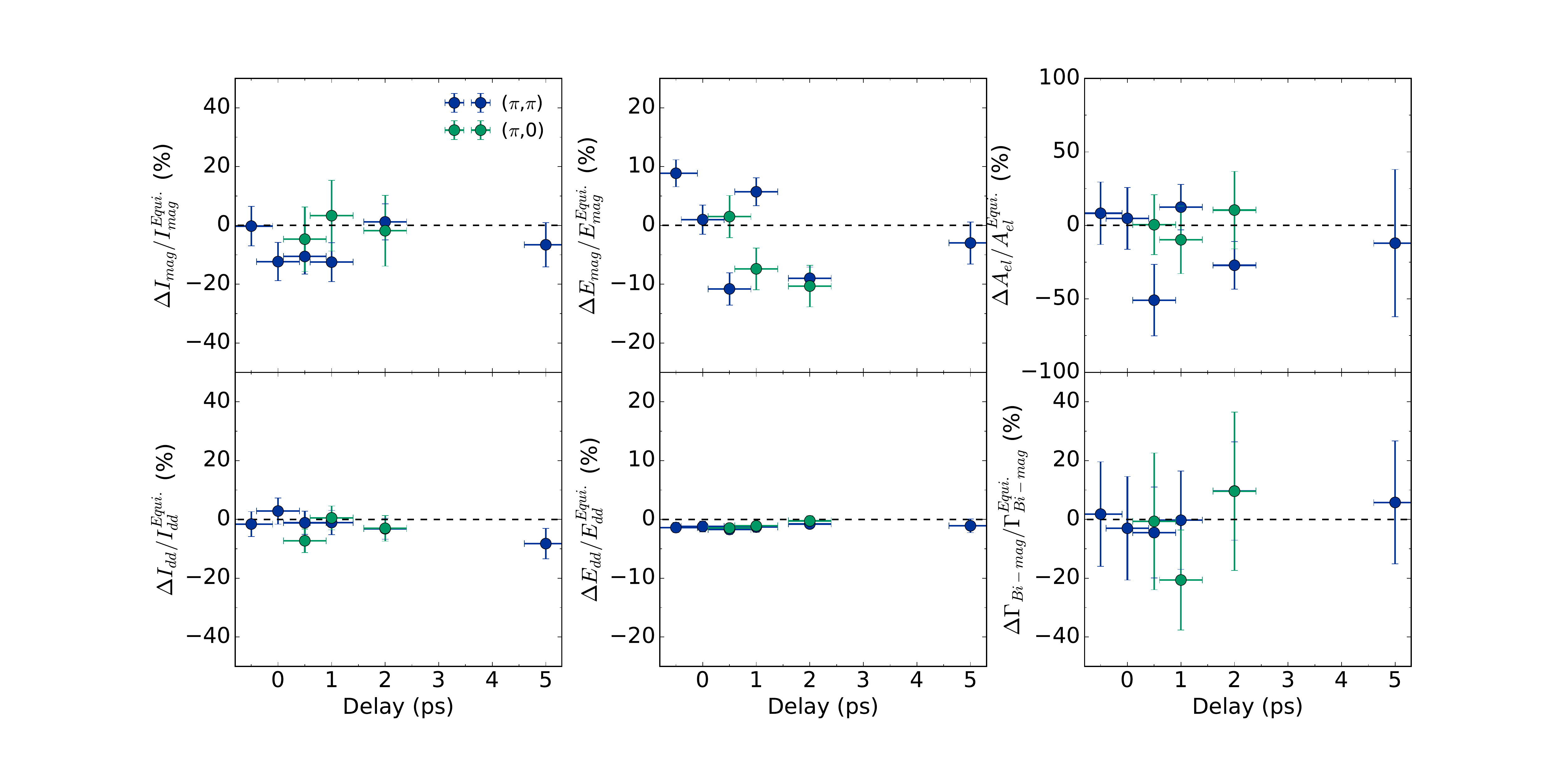}
\caption{Time evolution of RIXS fitting parameters. Relative time dependence of the integrated magnon, $I_\text{mag}$, and orbital, $I_\text{dd}$, intensity, the magnon and orbital position, $E_\text{mag}$ and $E_\text{dd}$, the amplitude of the elastic line, $A_\text{el}$, and the full-width at half-maximum of the broad magnetic feature $\Gamma_\text{Bi-mag}$ at higher energies.}
\label{SRIXS_analysis}
\end{figure*}

The \gls*{tr-RIXS} spectra in Fig.~2a and b of the main manuscript reveal an elastic line, a collective magnon at $\sim$100 and 150~meV for ($\pi$, $\pi$) and ($\pi$, 0), respectively, a magnon continuum at $\sim$250~meV and an orbital excitation at 680~meV. A sum of four Gaussian-shaped peaks was used to represent the energy-loss data. The center of the elastic line and its width were fixed in the analysis to 0~meV and the experimental resolution of 70 meV, respectively. Robust fits were obtained by further constraining the energy of the magnon continuum to 260~meV and its amplitude to the mean free value of 2.02 and 1.44 for ($\pi$, $\pi$) and ($\pi$, 0), respectively. All other parameters were varied freely. Figure~\ref{SRIXS_analysis} shows the relative time dependence of the integrated magnon, $I_\text{mag}$, and orbital, $I_\text{dd}$, intensity, the magnon and orbital position, $E_\text{mag}$ and $E_\text{dd}$, the amplitude of the elastic line, $A_\text{el}$, and the width of the broad magnon feature $\Gamma_\text{Bi-mag}$. Significant changes are observed only in the amplitude and width of the magnon and orbital excitation that are discussed in the main manuscript.

\section{SI Estimation of the X-ray and Laser penetration depth}
The optical penetration depth $\Lambda$ of a beam incident at angle $\alpha$ with respect to the sample surface is defined by \cite{Orfanidis2016}%
\begin{equation}
\Lambda = \frac{\lambda}{4\pi}\left[\Im\left(\sqrt{\tilde{N}^2-\cos^2(\alpha)}\right)\right]^{-1},
\label{pene}
\end{equation}
where $\Im$ signifies the imaginary part, $\lambda$ is radiation wavelength, and $\tilde{N} = n+ ik$ is the complex index of refraction. We define the penetration depth as the distance over which the beam intensity drops by $1/e$, and we remind the reader that $\tilde{N} = \sqrt{\tilde{\varepsilon}}$ where $\tilde{\varepsilon}$ is the relative permittivity. For the x-ray calculation, the dielectric constants below and above the Ir $L$-edge were calculated based on the crystal structure of Sr$_3$Ir$_2$O$_7$ using the xrayutilities Python package \cite{Kriegner2013}. The imaginary part of the index of refraction at the resonance was corrected considering the x-ray absorption white-line intensity \cite{Donnerer2018high}. At the Ir $L_3$-edge we find $\Lambda_{\text{x-ray}} = 162$, $214$ and $378$~nm for $\alpha = 1.6, 2.1$ and $3.7^\circ$, respectively. Using the reported optical conductivity results of Ref.~\cite{Ahn2016}, we find $\Lambda_{\text{laser}}$ $\approx 82$~nm for $\alpha \approx 20^\circ$.

\section{SI Spin-wave dispersion and estimation of N\'eel temperature}
The superexchange interactions in octahedrally coordinated 5$d^5$ iridates are strongly dependent on bond angles. In general, isotropic Heisenberg, anisotropic Heisenberg, Dzyaloshinskii-Moriya and Kitaev exchange couplings can be present. For the simplified case of exactly straight bonds, symmetric octahedra coordination and assuming that Coulomb repulsion, $U$, is much larger than Hund’s exchange, $J_\text{H}$, pseudo-dipolar and Kitaev interactions vanish by symmetry and the magnetic exchange becomes purely isotropic Heisenberg-like \cite{Jackeli2009}. Since Sr$_2$IrO$_4$ and Sr$_3$Ir$_2$O$_7$ possess iridium-oxygen-iridium bond angles that are close to 180$^\circ$ \cite{Jeffrey2016,Serrao2013, Hogan2016}, the dominant exchange term is Heisenberg-like, and the subdominant terms are anisotropic Heisenberg, Dzyaloshinskii-Moriya and Kitaev exchanges with the latter term generally considered negligible. This has been predicted in several theoretical works \cite{Takayama2016,Liu2019,Porras2019} and verified by RIXS and Raman scattering \cite{Kim2012, Kim2012_2, Gretarsson2015}. Thus, magnetism in Sr$_2$IrO$_4$ can be described by a Heisenberg Hamiltonian with in-plane isotropic coupling $J$, anisotropic coupling $\Gamma$, Dzyaloshinskii-Moriya interaction D and interlayer coupling $J_c$   \cite{Takayama2016}

\begin{equation}
   H_\text{214} = \sum_{\langle \vec{n}, \vec{m}\rangle}\big[J\vec{S}_{\vec{n}}\vec{S}_{\vec{m}}+\Gamma S_{\vec{n}}^zS_{\vec{m}}^z+D(-1)^{n_x+n_y}(S_{\vec{n}}^xS_{\vec{m}}^y-S_{\vec{n}}^yS_{\vec{m}}^x)\big] + \sum_{\langle\langle \vec{n},\vec{m}\rangle\rangle}J_\text{2}\vec{S}_{\vec{n}}\vec{S}_{\vec{m}}+ \sum_{\langle\langle\langle \vec{n},\vec{m}\rangle\rangle\rangle}J_\text{3}\vec{S}_{\vec{n}}\vec{S}_{\vec{m}}+\sum_{\vec{n},l}J_\text{c}\vec{S}_{\vec{n},l}\vec{S}_{\vec{n},l+1}.
    \label{Heisenberg214}
\end{equation}
Here, $\vec{n}$ = ($n_x$, $n_y$) are vectors to sites within the IrO$_2$ layers and $l$ is the layer index, defining $\vec{S}_{\vec{n},l}$ as the spin operator at site $\vec{n}$ of layer $l$. $\langle \vec{n},\vec{m}\rangle$, $\langle\langle \vec{n},\vec{m}\rangle\rangle$ and $\langle\langle\langle \vec{n},\vec{m}\rangle\rangle\rangle$ denote first, second and third nearest neighbors in the tetragonal plane and $J$, $J_\text{2}$ and $J_\text{3}$ are the corresponding isotropic interaction constants. We mention that some reported models also include an anisotropic, symmetric exchange term  in $H_\text{214}$ \cite{Liu2019,Porras2019}. This term is, however, much smaller than the interlayer coupling and can, thus, be neglected for our purposes.

In cases where the spins are aligned in the tetragonal plane, the nearest-neighbor interactions in Eq.~\ref{Heisenberg214} can be mapped onto an effective isotropic exchange Hamiltonian with interlayer coupling as \cite{Takayama2016}:

\begin{equation}
   H_\text{iso} = \sum_{\langle \vec{n},\vec{m}\rangle}\tilde{J}\vec{\bar{S}}_{\vec{n}}\vec{\bar{S}}_{\vec{m}}+\sum_{\vec{n},l}J_\text{c}\vec{\bar{S}}_{\vec{n},l}\vec{\bar{S}}_{\vec{n},l+1}.
    \label{iso}
\end{equation}
The transformation involves a rotation of the two sublattices in the tetragonal plane, where $\vec{\bar{S}}_{\vec{n},l}$ is the rotated spin operator and $\tilde{J}$ = $\sqrt{J^2+D^2}$. Since the transition temperature in the case of $J$ $\gg$ $J_c$ is dominated by the spin-wave velocity, we have neglected the second and third nearest neighbor couplings here. $T_\text{N}$ is defined as the temperature for which the expectation value $\langle \bar{S}^x_{\vec{n}}\rangle$ (calculated with linear spin-wave theory) vanishes. For $T_\text{N}$ $\ll$ $J/2$ the N\'eel temperature is given by \cite{Irkhin1997}
\begin{equation}
T_\text{N} = \frac{\pi\tilde{J}}{\text{log}\big(\frac{T_\text{N}^2}{4\tilde{J}J_\text{c}}\big)}.
    \label{Neel214}
\end{equation}
So if the interlayer coupling in Sr$_2$IrO$_4$ could be switched off, it would be predicted not to display magnetic order at any finite temperature.  Using $\tilde{J}$ = 50~meV to match a spin-wave energy of 200~meV at ($\pi$, 0), we can reproduce the experimentally observed $T_\text{N}$ = 285~K by assuming an interlayer coupling of 1.1~$\mu$eV. $T_\text{N}$ = 384.2~K is obtained when using the previously reported value $J_\text{c}$ = 15.9~$\mu$eV \cite{Takayama2016}.

As $c$-axis couplings gain particular importance in Sr$_3$Ir$_2$O$_7$ the Hamiltonian of Eq.~\ref{Heisenberg214} needs to be extended to 

\begin{align}
    H_\text{327} = &\sum_{\langle \vec{n},\vec{m}\rangle,l}\big[J\vec{S}_{\vec{n},l}\vec{S}_{\vec{m},l}+\Gamma S_{\vec{n},l}^zS_{\vec{m},l}^z\big] + \sum_{\langle \vec{n},\vec{m}\rangle,l} D(-1)^{n_x+n_y+l}(S_{\vec{n},l}^xS_{\vec{m},l}^y-S_{\vec{n},l}^yS_{\vec{m},l}^x)\nonumber\\
    &\sum_{\vec{n}}\big[J_\text{c}\vec{S}_{\vec{n},1}\vec{S}_{\vec{n},2}+\Gamma_\text{c}S_{\vec{n},1}^zS_{\vec{n},2}^z\big] + \sum_{\vec{n}} D_\text{c}(-1)^{n_x+n_y}(S_{\vec{n},1}^xS_{\vec{n},2}^y-S_{\vec{n},1}^yS_{\vec{n},2}^x)
    \label{Heisenberg327}
\end{align}
with the additional terms $\Gamma_\text{c}$ and $D_\text{c}$, which couple the planes within a bilayer. This model neglects coupling between the bilayers. The N\'eel temperature of a collinear antiferromagnetic spin alignment is found from the numerical solution of $\langle S^z_{\vec{n},l}\rangle$ = 0 with

\begin{equation}
\langle S^z_{\vec{n},1}\rangle = 1-\frac{1}{16\pi^2}\int\bigg[\frac{B_{\vec{q},+}}{E_+(\vec{q})}\coth\big(\beta E_+(\vec{q})\big) + \frac{B_{\vec{q},-}}{E_-(\vec{q})}\coth\big(\beta E_-(\vec{q})\big)\bigg]\text{dq$_\text{x}$dq$_\text{y}$}.
\end{equation}
$E_\pm(\vec{q})$ = $\sqrt{B_{\vec{q},\pm}^2-|C_{\vec{q},\pm}|^2}$ define the two magnon branches with
\begin{align*} 
&\beta=1/(k_\text{B}T_N)\\
&B_{\vec{\vec{q}},\pm} = \frac{1}{2}(8J + 8\Gamma + J_c + \Gamma_c)-4J_2(1-\cos(q_x)\cos(q_y))-4J_3(1-\gamma_{2\vec{q}})-2J_{2c}(1\mp \gamma_{\vec{q}})\\
&C_{\vec{q},\pm} = \frac{1}{2}(8J\gamma_{\vec{q}} \pm J_c) - \frac{i}{2}(8D\gamma_{\vec{q}} \pm D_c)\\
&\gamma_{\vec{q}} = \frac{1}{2}(\cos(q_x)+\cos(q_y))
\end{align*}
and $k_\text{B}$ = 8.617$\cdot$10$^{-2}$~meV/K the Boltzmann constant. A reasonable N\'eel temperature $T_\text{N}$ = 118~K, which is within a factor of three of the experimental value, is obtained for $J$ = 46.6~meV,  $J_\text{c}$ = 25.2~meV,  $J_\text{2}$ = 5.95~meV, $J_\text{3}$ = 7.3~meV,  $J_\text{2c}$ = 6.2~meV, $\Gamma$ = 2.2~meV, $\Gamma_\text{c}$ = 34.3~meV, $D$ = 12.25~meV and $D_\text{c}$ = 28.1~meV. The analysis shows that magnetic order perpendicular to the tetragonal plane can be stabilized even in the absence of long-range $c$-axis coupling. Furthermore, in Sr$_3$Ir$_2$O$_7$ we do not expect the same extreme anisotropy between the $ab$-plane and $c$-axis correlations as in Sr$_2$IrO$_4$.  
\bibliography{refs}